\documentclass[12pt,twocolumn]{emulateapj}

\usepackage{color}
\usepackage[usenames,dvipsnames]{xcolor}
\usepackage[colorlinks,urlcolor=blue,citecolor=blue,linkcolor=blue]{hyperref}
\usepackage[multidot]{grffile} 
\usepackage{amsmath}
\usepackage{array}
\usepackage{xfrac}
\usepackage{multirow}

\slugcomment{}

\begin{document} 

\title{APOGEE DR14/DR15 Abundances in the Inner Milky Way} 
\shorttitle{Inner Milky Way Abundances}

\author{
G.~Zasowski\altaffilmark{1}, 
M.~Schultheis\altaffilmark{2},
S.~Hasselquist\altaffilmark{3},
K.~Cunha\altaffilmark{4,5}, 
J.~Sobeck\altaffilmark{6}, \\
J.~A.~Johnson\altaffilmark{7},
A.~Rojas-Arriagada\altaffilmark{8,9},
S.~R.~Majewski\altaffilmark{10},
B.~H.~Andrews\altaffilmark{11}, 
H.~J\"{o}nsson\altaffilmark{12}, \\
T.~C.~Beers\altaffilmark{13}, 
S.~D.~Chojnowski\altaffilmark{3},
P.~M.~Frinchaboy\altaffilmark{14}, 
J.~A.~Holtzman\altaffilmark{3}, \\
D.~Minniti\altaffilmark{8}, 
D.~L.~Nidever\altaffilmark{15,16}, 
C.~Nitschelm\altaffilmark{17} 
}
\shortauthors{Zasowski et al.}

\altaffiltext{1}{Department of Physics \& Astronomy, University of Utah, Salt Lake City, UT, 84112, USA; gail.zasowski@gmail.com}
\altaffiltext{2}{Laboratoire Lagrange, Universit\'{e} C\^{o}te d'Azur, Observatoire de la C\^{o}te d'Azur, 06304, Nice, France}
\altaffiltext{3}{Department of Astronomy, New Mexico State University, Las Cruces, NM, 88001, USA}
\altaffiltext{4}{Steward Observatory, The University of Arizona, Tucson, AZ, 85719, USA}
\altaffiltext{5}{Observat\'{o}rio Nacional, 20921-400 So Crist\'{o}vao, Rio de Janeiro, RJ, Brazil}
\altaffiltext{6}{Department of Astronomy, University of Washington, Seattle, WA, 98195, USA}
\altaffiltext{7}{Department of Astronomy, The Ohio State University, Columbus, OH, 43210, USA}
\altaffiltext{8}{Instituto de Astrof\'{i}sica, Pontificia Universidad Cat\'olica de Chile, Santiago, Chile}
\altaffiltext{9}{Millennium Institute of Astrophysics, 782-0436 Macul, Santiago, Chile}
\altaffiltext{10}{Department of Astronomy, University of Virginia, Charlottesville, VA, 22904, USA}
\altaffiltext{11}{Department of Physics and Astronomy \& the Pittsburgh Particle Physics, Astrophysics and Cosmology Center (PITT PACC), University of Pittsburgh, Pittsburgh, PA, 15260, USA}
\altaffiltext{12}{Lund Observatory, Department of Astronomy and Theoretical Physics, Lund University, SE-221 00 Lund, Sweden}
\altaffiltext{13}{Department of Physics, University of Notre Dame, \& JINA Center for the Evolution of the Elements, Notre Dame, IN, 46556 USA}
\altaffiltext{14}{Department of Physics \& Astronomy, Texas Christian University, Fort Worth, TX, 76129, USA}
\altaffiltext{15}{National Optical Astronomy Observatory, Tucson, AZ, 85719, USA}
\altaffiltext{16}{Department of Physics, Montana State University, ozeman, MT, 59717, USA}
\altaffiltext{17}{Unidad de Astronom\'{i}a, Universidad de Antofagasta, Antofagasta 1270300, Chile}

\begin{abstract}
We present an overview of the distributions of 11 elemental abundances in the Milky Way's inner regions, as traced by APOGEE stars released as part of SDSS Data Release~14/15 (DR14/DR15), including O, Mg, Si, Ca, Cr, Mn, Co, Ni, Na, Al, and K.  This sample spans $\sim$4000 stars with $R_{\rm GC} \le 4.0$~kpc, enabling the most comprehensive study to date of these abundances and their variations within the innermost few kiloparsecs of the Milky Way.  
We describe the observed abundance patterns ([X/Fe]--[Fe/H]), compare to previous literature results and to patterns in stars at the solar Galactocentric radius ($R_{\rm GC}$), and discuss possible trends with DR14/DR15 effective temperatures.
We find that the position of the [Mg/Fe]--[Fe/H] ``knee'' is nearly constant with $R_{\rm GC}$, indicating a well-mixed star-forming medium or high levels of radial migration in the early inner Galaxy.  We quantify the linear correlation between pairs of elements in different subsamples of stars and find that these relationships vary; some abundance correlations are very similar between the $\alpha$-rich and $\alpha$-poor stars, but others differ significantly, suggesting variations in the metallicity dependencies of certain supernova yields.  These empirical trends will form the basis for more detailed future explorations and for the refinement of model comparison metrics.  That the inner Milky Way abundances appear dominated by a single chemical evolutionary track and that they extend to such high metallicities underscore the unique importance of this part of the Galaxy for constraining the ingredients of chemical evolution modeling and for improving our understanding of the evolution of the Galaxy as a whole.
\end{abstract}

\section{Introduction}
The Milky Way (MW) galaxy is often described as the best local laboratory for studying galaxy evolution (not just Galaxy evolution), and nowhere is this more true than in studies of its bulge, bar, and inner disk regions.  Individual stars in comparable regions of other large galaxy systems are thus far unresolvable, especially for spectroscopy, so the MW remains the only system with which to study the chemical diversity and chemodynamical relationships of stellar populations in those parts of galaxies where most stars live.

The MW's inner region contains a bar with a boxy-peanut shape \citep[e.g.,][]{Wegg_2013_RCbulge} and an X-shaped flare \citep[][but see also \citealt{Han_2018_noXshape}]{McWilliam_10_xshapedbulge,Nataf_2010_dualbulgeRC,Saito_2011_Xshapedbulge,Ness_2016_Xshapedbulge}, the inner disk and disk-bar transition, the inner part of the halo, and possibly a merger-dominated classical bulge of unknown mass and size \citep[e.g.,][]{Shen_10_purediskbulge,Nataf_2017_bulgeformation}.  For simplicity, in this paper we will refer to this entire complex as ``the bulge'' or ``the inner MW'' (see \S\ref{sec:sel_criteria}).  

The MW's bulge is a dense environment with a very long ``star accumulation'' history \citep[including the accretion of stars and in~situ formation;][]{Nataf_2017_bulgeformation,Barbuy_2018_bulgereview} that may cause it to differ in chemistry and star-formation history markers from populations farther out in the disk. These populations at larger Galactocentric radii ($R_{\rm GC}$), particularly in the solar neighborhood, are the ones upon which most of our nucleosynthesis models are tested.  We must identify and understand any differences with the chemical patterns of the bulge to expand and validate these models and to compare with integrated-light abundances in the central regions of other galaxies. 

Historically, our characterization of the chemistry of the inner MW is drawn from numerous, diverse samples of dozens to a few thousand stars each, typically located in pencil-beam sightlines probing lower-extinction windows \citep[for some early examples, see][]{McWilliam_94_KgiantsBW,Rich_05_MgiantsBW,Fulbright_2006_baadewinchem,Cunha_06_bulgespectra}.  Despite the heterogeneous nature of these samples --- in terms of facilities, wavelength range, spectral resolution, and so on --- these data represent an incredible leap over what was known even two decades ago. The bulge is a predominantly old \citep[e.g., $t \gtrsim 8-10$~Gyr;][]{Nataf_2015_bulgeSFH}, intermediate- to high-metallicity \citep[${\rm -0.75 \lesssim [Fe/H] \lesssim +0.5}$; e.g.,][]{Ness_2016_bulgeMDF} population with chemical patterns, particularly in the $\alpha$-elements and in some heavy Fe-peak elements, very similar to the $\alpha$-enhanced disk\footnote{We adopt a chemistry-based terminology in lieu of the ``thin vs.\ thick disk'' paradigm to avoid confusion with morphological and kinematical distinctions \citep[e.g.,][]{Martig_2016_thickdiskages}.  For example, canonical ``thin disk'' stars (${\rm [Fe/H] \lesssim -0.25}$, ${\rm [\alpha/Fe] \lesssim +0.1}$) are frequently found at high $|Z_{\rm GC}|$ \citep[e.g.,][]{Boeche_2013_RAVEchemodynamics,Nidever_2014_MWchemevolution,hayden15}. While this particular example is most common in the outer disk \citep[perhaps due to disk flaring; e.g.,][]{Minchev_2015_thickdisks,Mackereth_2017_diskagemetallicity}, it highlights the need for clear definitions.} at larger $R_{\rm GC}$ \citep[e.g.,][]{Zoccali_2003_ageFeHbulge,Zoccali_06_oxygeninbulge,Clarkson_2008_bulgePMs,Alves-Brito_2010_bulgethickdisk,Bensby_2013_bulgeages,Gonzalez_2015_gibs,Bensby_2017_bulgedwarfsSGs,RojasArriagada_2017_GESbulge}.
In addition to this global picture, observations of a significant younger population at high metallicity have also been reported \citep[e.g., $t<8$~Gyr for more than a third of stars with ${\rm [Fe/H] > 0}$;][]{Bensby_2017_bulgedwarfsSGs}.  The metallicity distribution also has a small but well-measured tail extending lower than ${\rm [Fe/H] < -1}$ \citep[e.g.,][]{Howes_2014_GESmpbulge,Howes_2015_bulgeEMPstars,Koch_2016_MPbulge,Kunder_2016_bulgeRRLbrava,ContrerasRamos_vvvRRLbulge}.

Large spectroscopic surveys, especially at red-optical or infrared (IR) wavelengths, offer immense power to expand and refine our mapping of the inner Galaxy's mean chemical properties and our understanding of its more subtle nuances that require large statistical samples --- e.g., the detailed chemical substructure and abundance patterns of different families of stars or the discovery of rarer populations. At red-optical wavelengths, analysis of the ARGOS \citep[Abundances and Radial velocity Galactic Origins Survey;][]{Freeman_2013_argos}, GIBS \citep[GIRAFFE Inner Bulge Survey;][]{Zoccali_2014_GIBS1}, and Gaia-ESO \citep{Gilmore_2012_GaiaESOspectro,Randich_2013_GaiaESOspectro} surveys have provided insight into chemo-dynamical subpopulations and patterns that span several degrees of the inner MW \citep[e.g.,][]{Ness_2013_argoskinematics,Zoccali_2017_GIBS3,RojasArriagada_2017_GESbulge}.

The APOGEE survey \citep[\S\ref{sec:apogee};][]{Majewski_2017_apogee1} provides a particularly powerful dataset for this type of exploration due to its $H$-band sensitivity, extensive field of view, and large, homogeneous, statistical sample of stars. 

For example, \citet{GarciaPerez_2013_MPbulge} were able to study the abundances of rare, very metal-poor stars in the inner MW using data from APOGEE (\S\ref{sec:apogee}).
\citet{Schiavon_2017_APOGEEbulgeGCs} explored the presence of multiple stellar populations in inner Galaxy globular clusters ($R_{\rm GC} < 2.2$~kpc), while \citet{Schiavon_2017_Nichbulge} discovered a corresponding field star population in the inner Galaxy with chemical abundances (C, N, Al) similar to that of globular clusters, which were also discussed by \citet{FernandezTrincado_2017_2ndGenGCs}. These N-enhanced stars could be either former members of dissolved globular clusters or by-products of similar chemical enrichment by the first generations of stars formed in the inner Milky Way. 

\citet{GarciaPerez_2018_dr12mdf} provided APOGEE's first large study of the bulge metallicity distribution function (MDF). The decomposition of the MDFs in different sightlines suggests approximately four different metallicity components (peaking between ${\rm [Fe/H]} \sim -0.8$ and $\sim +0.3$) with varying relative strengths across the bulge. \citet{Schultheis_2017_apogeeBW} showed that the APOGEE MDF in Baade's Window \citep[using DR13 data;][]{SDSS_2016_dr13} agreed extremely well with that of {\it Gaia}-ESO \citep{rojas2014}, despite the surveys' different selection functions. 

These metallicity ``components'' may be associated with populations in the bar, the disk, or the inner halo \citep[e.g.,][]{Ness_2016_bulgeMDF}. Efforts are underway to test whether these associations are robust.  For example, \citet{Ness_2016_apogeekinematics} and \citet{Zasowski_2016_apogeebulge} used APOGEE data to study the relationships between stellar metallicity and radial velocity distribution moments  in the inner $\sim$4~kpc of the MW.
\citet{Portail_2017_chemodynbulgemodel} employed Made-to-Measure modeling to derive orbital dynamics for stars in different metallicity components.  In addition, by using $N$-body simulations, \citet{Fragkoudi_2018_bulgeNbody} were able to reproduce the APOGEE MDF in the inner MW using a multi-component disk that evolves secularly to form a bar and  a boxy bulge \citep[see also][]{GarciaPerez_2018_dr12mdf}. 

In this paper, we complement these efforts by providing an empirical description of several of the elemental abundance patterns of inner MW stars provided in the latest SDSS data release.  We describe the dataset and sample selection in \S\ref{sec:sample}.  In \S\ref{sec:xfe_vs_feh}, we discuss the properties of each [X/Fe]--[Fe/H] plane, including aspects that are likely to be ``real'' and those that are likely to be artifacts of the analysis process.  \S\ref{sec:sn_comparison} contains a comparison between the abundance patterns of the bulge sample and a matched set of stars near the solar radius, \S\ref{sec:knee} presents a measurement of the [Fe/H] reached at the onset of Type~Ia supernovae at different radii in the inner MW, and \S\ref{sec:correlations} describes the correlations between pairs of elements in different subsamples of the bulge population.  Our findings are summarized in \S\ref{sec:summary}. 

\section{Sample}
\label{sec:sample}

\subsection{APOGEE Data}
\label{sec:apogee}

We use data from the Sloan Digital Sky Survey (SDSS) Data Release 14 \citep[DR14;][]{Abolfathi_2018_SDSSDR14}, which includes spectra, stellar parameters, and stellar abundances from the APOGEE-1 and APOGEE-2 surveys \citep{Majewski_2017_apogee1}.  These data are identical to those appearing in SDSS DR15 (Aguado et al., submitted). APOGEE-1 was a component of the SDSS-III \citep{Eisenstein_11_sdss3overview}, and APOGEE-2 is part of the SDSS-IV \citep{Blanton_2017_sdss4}.  Both APOGEE-1 and APOGEE-2 North projects utilize the 2.5-m Sloan Telescope at Apache Point Observatory \citep{Gunn_2006_sloantelescope}, coupled to a 300-fiber, high-resolution ($R \sim 22,000), $ $H$-band spectrograph \citep{Wilson_2012_apogee}.  A second spectrograph is currently taking observations as part of APOGEE-2 South on the 2.5-m du~Pont Telescope at Las Campanas Observatory; inner MW data from this southern survey component will be available in future data releases.

The primary APOGEE sample comprises spectra of red giant stars in the magnitude range $7 \lesssim H \lesssim 13.8$, reduced with a custom pipeline.  For details of the target selection in APOGEE-1 and -2, see \citet[][]{zasowski2013} and \citet{Zasowski_2017_apogee2targeting}, respectively.  \citet{Nidever_2015_apogeereduction} describe the reduction and radial velocity (RV) measurement pipelines, and \citet{GarciaPerez_2016_aspcap} describe the APOGEE Stellar Parameter and Abundances Pipeline (ASPCAP).  

Throughout this paper, we use the DR14/DR15 calibrated abundances.
Details of the data calibration and available data products can be found in \citet{meszaros2013}, \citet{holtzman2015}, and \citet{Holtzman_2018_dr13dr14apogee}.
The APOGEE abundances are calibrated to remove systematic uncertainties as much as possible; using comparisons to optical-based abundances for the same stars, \citet{Jonsson_2018_dr13dr14abundances} show that while some systematics are likely to remain for certain elements (e.g., N, K, V), this procedure is generally effective.  The individual, random abundance uncertainties are estimated by deriving an empirical fit for the abundance scatter within stellar clusters. This fit is a function of $T_{\rm eff}$, [M/H], and SNR; it does not explicitly include systematics, but the comparison in \citet{Jonsson_2018_dr13dr14abundances} gives us reason to believe these are generally small. For the majority of elements in the majority of stars, the quoted uncertainties are $< 0.1$~dex.

\subsection{Selection Criteria}
\label{sec:sel_criteria}

Starting with the full DR14/DR15 catalog, we first removed stars flagged as not part of the primary red giant sample (i.e., ancillary targets and pre-selected Galactic Center supergiants).  We also removed stars without reliable effective temperatures, surface gravities, and metallicities derived by ASPCAP, since without well-determined parameters, any derived elemental abundances are highly uncertain.  For these culls, we required that the ASPCAPFLAG bits 19, 20, and 23 and STARFLAG bit 9 not be set (the METALS\_BAD, ALPHAFE\_BAD, STAR\_BAD, and PERSIST\_HIGH flags\footnote{\url{http://www.sdss.org/dr14/algorithms/bitmasks/}}, respectively).  We also removed stars outside the ranges $3600 \leq T_{\rm eff} \leq 4500$~K and $-0.75 \leq \log{g} \leq 3.5$ to ensure a sample of similarly evolved, inner MW giant stars with the most reliable parameters.  Abundances with uncertainties larger than 0.15~dex are eliminated from the analyses described below.

\begin{figure}
\begin{center}
\includegraphics[angle=0,trim=0in 0.5in 0in 0.95in, clip, width=0.5\textwidth]{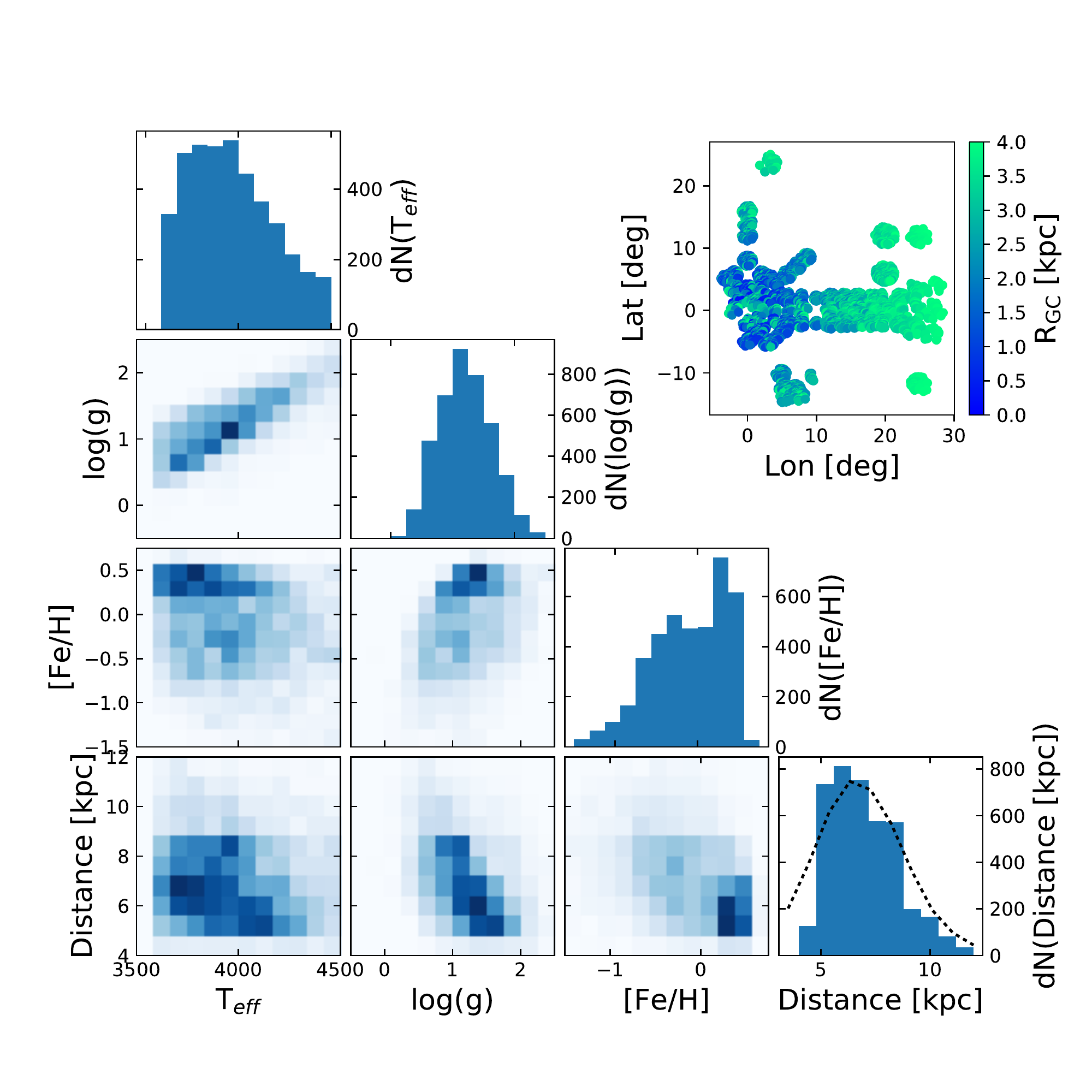} 
\caption{
Joint parameter ($T_{\rm eff}$, $\log{g}$, [Fe/H]) and heliocentric distance distributions for the inner Galaxy sample described in \S\ref{sec:sel_criteria}.  The upper-right inset contains the Galactic $(\ell,b)$ distribution of stars, colored by Galactocentric distance $R_{\rm GC}$ as indicated.  The dotted line in the distance distribution panel (bottom right) repeats the data in the blue histogram, blurred by the individual distance uncertainties.
}
\label{fig:sample_params}
\end{center}
\end{figure}

Heliocentric distances for the stars in DR14/DR15 have been calculated by multiple groups, including \citet{Wang_2016_APOGEEdistances}, \citet{Schultheis_2017_apogeeBW}, and \citet[][using an updated version of \citealt{Santiago_2016_apogeedistances}'s code]{Queiroz_2018_starhorse}\footnote{\url{http://www.sdss.org/dr14/data_access/value-added-catalogs/?vac_id=apogee-dr14-based-distance-estimations}}.  A detailed comparison of these distance sets is beyond the scope of this paper, but in general the agreement is good.  We adopt the distances of \citet{Queiroz_2018_starhorse}, which incorporate the recent extinction law of \citet{Schlafly_2016_OIRextinctionlaw} and newer Galactic structural priors from \citet{robin2012} and \citet{BlandHawthorn_2016_theMWproperties}.
From this catalog, we used the median posterior distance and the posterior distance standard deviation for the distance and distance uncertainty, respectively; we confirmed that the additional PDF percentiles reported are consistent with gaussian PDFs, so we treat the uncertainties as gaussian throughout this paper (e.g., in Figure~\ref{fig:sample_params} and \S\ref{sec:knee}). We note that the qualitative conclusions in this paper are independent of which distance set is used.  The variance in sample size when using different sets for the distance limits below is $\sim$20\%, but we find no systematic trend (in terms of stellar parameters) in the stars that meet our requirements using one particular distance set.

Using a solar distance of $R_0 = 8.3$~kpc \citep[e.g.,][]{Chatzopoulos_2015_GCstarcluster}, we apply a Galactocentric distance limit on the stellar sample of $R_{\rm GC} \leq 4.0$~kpc, a range that includes the so-called long or thin bar \citep[e.g.,][]{Wegg_2015_RClongbar}, and a fractional $R_{\rm GC}$ uncertainty limit of $\le$40\%.  The resulting sample comprises 4058 stars with the stellar parameter and heliocentric distance distributions shown in Figure~\ref{fig:sample_params}. The dotted line in the bottom right panel shows the distance distribution as a summation of $N$ gaussians, one for each star in the sample, where each gaussian is centered at the star's calculated distance and broadened by the distance uncertainty (the median fractional uncertainty in the sample is 12.2\%, with a standard deviation of 4\%).  This distribution is broader than the simple histogram, unsurprisingly, but the similar shape suggests that the true distance distribution is not significantly different than what is assumed in the following analyses by using the quoted values.

This selected spatial region includes stars in the bar, the inner halo, the $\alpha$-enhanced and $\alpha$-solar components of the disk, and any other structural components in the center of the MW. We do not attempt to disentangle these components kinematically or chemically; 
our goal here is to describe the abundance patterns of the sum total of the populations residing in the inner regions of the Galaxy.

\section{Discussion}

\subsection{[X/Fe] vs. [Fe/H]}
\label{sec:xfe_vs_feh}

The APOGEE DR14/DR15 release includes elemental abundances for 22 elements: C, N, O, Na, Mg, Al, Si, P, S, K, Ca, Ti, V, Cr, Mn, Co, Ni, Cu, Ge, Rb, Yb\footnote{Erroneously labeled ``Y'' in the DR14/DR15 data products.  This will be fixed in future data releases.}, and Nd.
However, as discussed in \citet{Jonsson_2018_dr13dr14abundances}, several of these abundances show unexpectedly large scatter, due primarily to the impact of weak lines and/or line blending. Here, we restrict our discussion to 11 elements: O, Mg, Si, Ca, Cr, Mn, Co, Ni, Na, Al, and K, a selection motivated by the removal of species dominated by scatter unreflected in the uncertainties and of those whose surface abundances undergo substantial evolution during a stellar lifetime (e.g., C and N).  
This set of elements largely corresponds to the abundances assessed by \citet{Jonsson_2018_dr13dr14abundances} to have small systematic and random differences when compared to abundances from optical spectroscopy.  However, we also include some with large scatter or other unusual behavior (e.g., K and Co), because these abundance uncertainties seem to be mostly systematic in nature \citep{Jonsson_2018_dr13dr14abundances} and can be used in a differential comparison between the inner MW sample and a disk sample of similar stars (below) that should display the same systematics.

Throughout the discussion below, we refer to several literature studies as well as two primary APOGEE references: the mean trends observed in literature datasets compared with APOGEE DR13 abundances in Baade's Window analyzed in \citet{Schultheis_2017_apogeeBW}, and the detailed assessment of APOGEE DR13--DR15 abundance accuracy, precision, and consistency with literature values provided by \citet{Jonsson_2018_dr13dr14abundances}.
We also refer to a useful stellar comparison sample: stars at approximately the solar circle (solar radius; SR) selected with quality criteria identical to those described in \S\ref{sec:sel_criteria}, but with $6 < R_{\rm GC} < 10$~kpc and $|Z_{\rm GC}| < 1.25$~kpc.  The SR stars are then subsampled to have a very similar joint $T_{\rm eff}$--[Fe/H] distribution as the inner Galaxy sample; this facilitates direct comparison of the abundance patterns without the impact of the disparate metallicity distributions and differently sampled RGBs in the two Galactic regions (Appendix~\ref{sec:app_snsample}).

\begin{figure*}[!hptb]
\begin{center}
\includegraphics[angle=0,trim=0.2in 1.8in 0in 1.8in, clip, width=\textwidth]{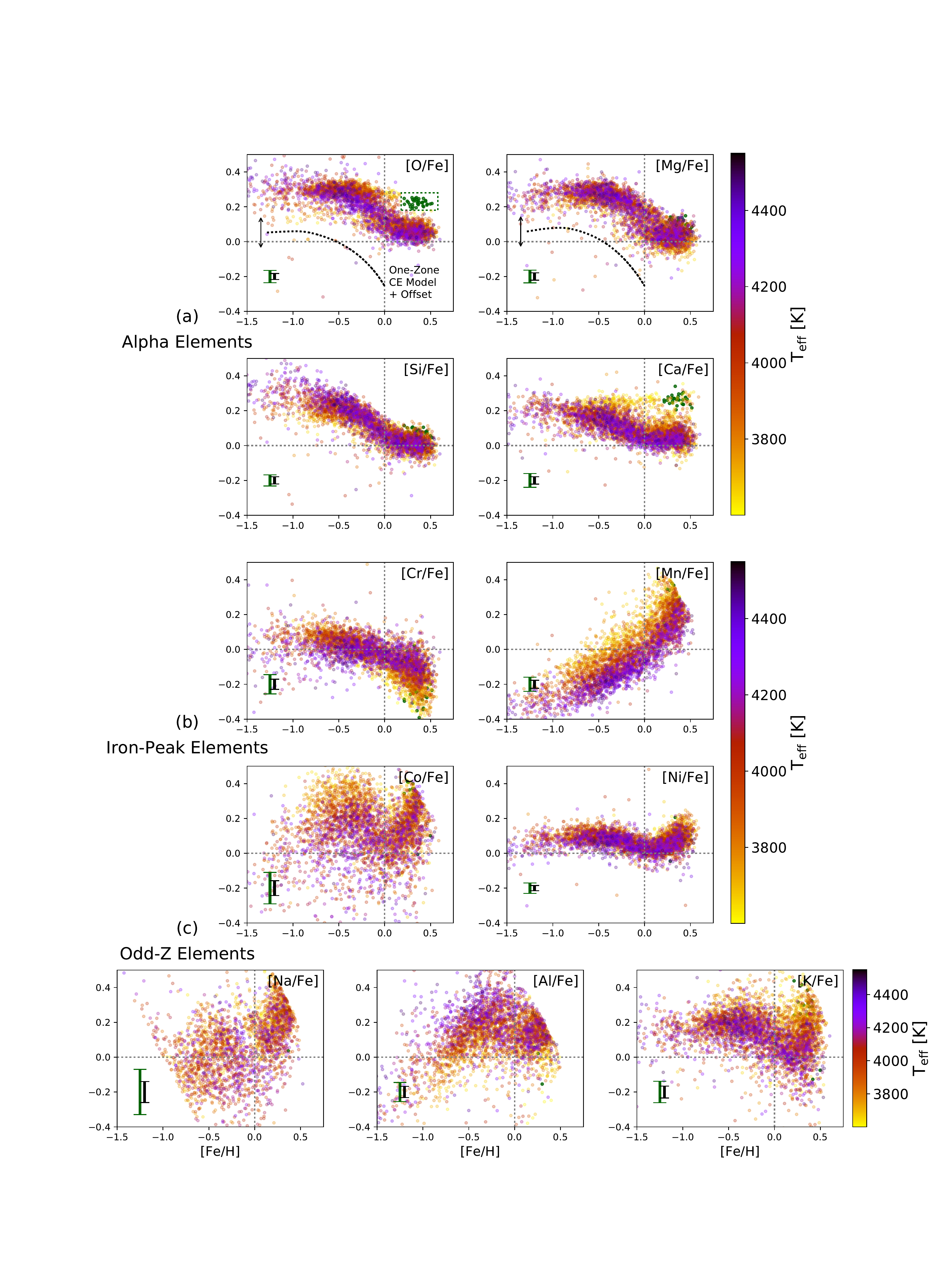} 
\caption{
[Fe/H]--[X/Fe] distributions for the elemental abundances in this paper: (a) $\alpha$ elements (\S\ref{sec:alpha_elements}), (b) iron-peak elements (\S\ref{sec:fepeak_elements}), and (c) odd-Z elements (\S\ref{sec:oddz_elements}).  The color of each point indicates the star's effective temperature; the green points are the ``high-O'' stars excluded from further analysis (\S\ref{sec:alpha_elements} and Appendix~\ref{sec:app_highO}).  The black (green) error bar in the lower left corner of each panel shows the median (95th percentile) uncertainty in each abundance.  Horizontal and vertical gray dotted lines indicate solar values, and the black dotted lines are one-zone chemical evolution sequences as described in the text (the vertical arrows emphasize that their vertical shifts are arbitrary).
}
\label{fig:abundances_grid}
\end{center}
\end{figure*}

\begin{figure*}
\begin{center}
\includegraphics[angle=0,trim=0.2in 1.8in 0in 1.8in, clip, width=\textwidth]{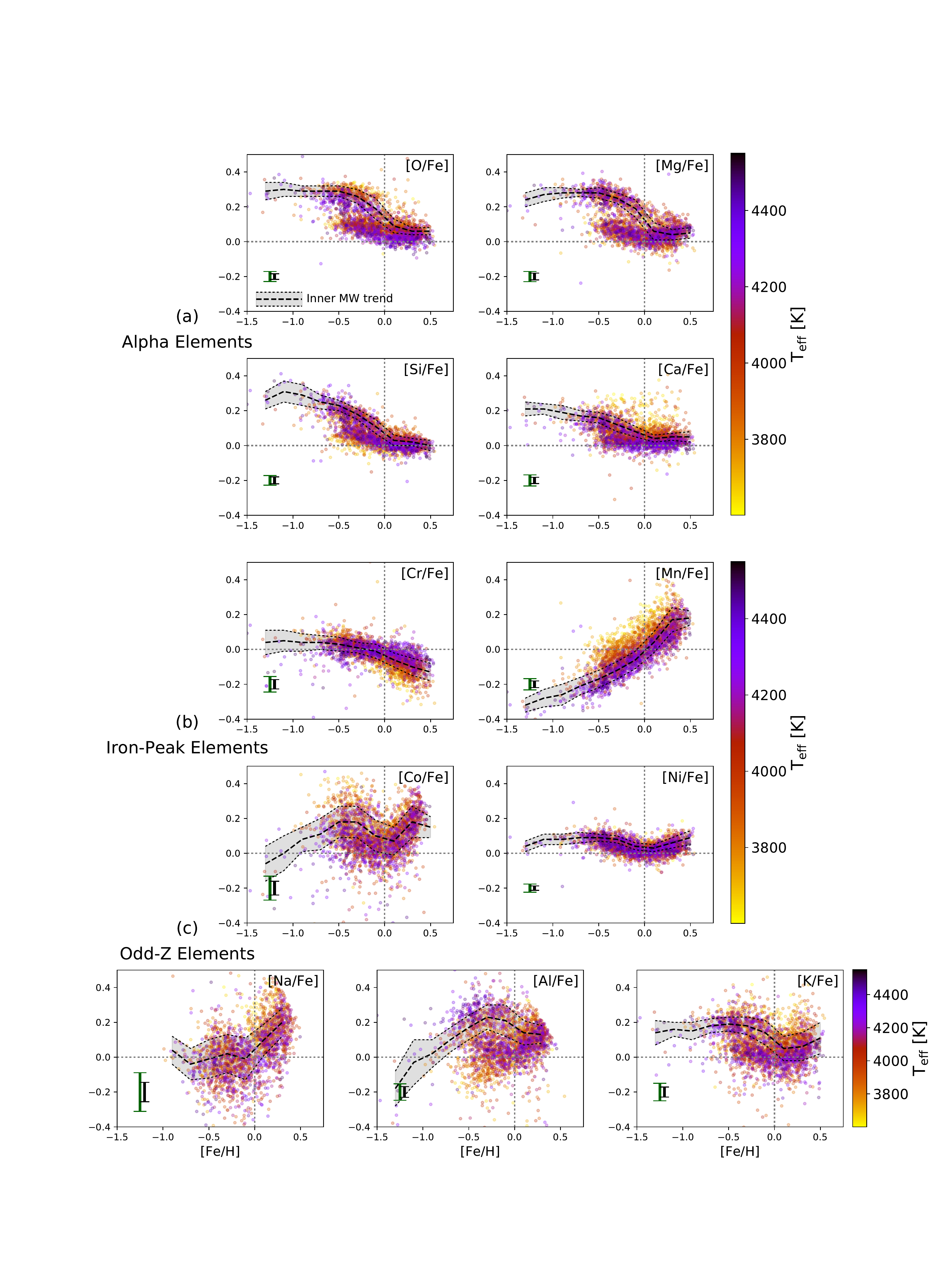} 
\caption{
[Fe/H]--[X/Fe] distributions for the elemental abundances in the solar radius comparison sample (\S\ref{sec:sel_criteria}).  As in Figure~\ref{fig:abundances_grid}, the color of each point indicates the star's effective temperature, and the error bars in the lower left corner of each panel show the median and 95th percentile uncertainties in each abundance.
The dashed line and gray swath in each panel indicates the median trend and $\pm$1$\times$ the median absolute deviation of the abundances of inner MW stars with $T_{\rm eff} \ge 3800$~K in Figure~\ref{fig:abundances_grid}.
}
\label{fig:abundances_grid_sn}
\end{center}
\end{figure*}

Figure~\ref{fig:abundances_grid} contains the [X/Fe]--[Fe/H] distributions for our chosen elements in the inner MW stars. Figure~\ref{fig:abundances_grid_sn} shows the comparable distributions for the SR sample, as well as the median and $\pm$1$\sigma$ trends of the inner MW stars from Figure~\ref{fig:abundances_grid} for comparison. We discuss each of these abundance distributions in \S\S\ref{sec:alpha_elements}--\ref{sec:oddz_elements} and provide some summary statistics of the distributions in Table~\ref{tab:abundances_properties}.
For a detailed comparison of overall ASPCAP abundances to various literature sources, see \citet{Jonsson_2018_dr13dr14abundances}. Here we focus on the observed patterns as released in DR14/DR15 and on literature values from the inner Galaxy, with some comparisons to chemical enrichment yield patterns from \citet{Andrews_2017_flexCE} that use nucleosynthetic yields compiled from \citet{woosley1995}, \citet{Iwamoto_1999_SNIayields}, \citet{Chieffi_2004_SNyields}, \cite{Limongi_2006_CCSNyields}, and \citet{Karakas_2010_AGByields}.

Figure~\ref{fig:abundances_grid} also contains one-zone chemical evolution sequences for the [O/Fe] and [Mg/Fe] abundances (black dotted lines).  These sequences have been shifted vertically by an arbitrary amount because we want to emphasize that these are not fits to the data, but rather that the similarity in shape demonstrate that the $\alpha$-abundances, at least, appear dominated by a simple evolutionary track (in contrast to the SR distributions in Figure~\ref{fig:abundances_grid_sn}).  These tracks, computed with the flexCE code\footnote{\url{https://github.com/bretthandrews/flexCE}} of \citet{Andrews_2017_flexCE}, have the same parameters as the fiducial model in that paper, with the exception of the outflow mass-loading factor (here, $\eta = 2.0$). A comprehensive fitting and comparison to a wider range of chemical evolution models is deferred to future work; here we provide empirical descriptions of the data and its internal relationships, which will serve as observational constraints to these chemical evolution models.

\begin{table*}[!hptb]
\begin{center}
\begin{tabular}{c| ccc | ccc | ccc}
 \hline
 \hline
  \multirow{2}{*}{Element}  & \multicolumn{3}{c}{Median Abundance (MAD)} & \multicolumn{3}{c}{Median Uncertainty} & \multicolumn{3}{c}{Trend as [Fe/H] Increases} \\
  & $\rm{[Fe/H]} < -0.8$ & $-0.8 \le \rm{[Fe/H]} < 0$ & $\rm{[Fe/H]} \ge 0$ & $<-0.8$ & $-0.8$ to 0 & $\ge 0$ & $<-0.8$ & $-0.8$ to 0 & $\ge 0$  \\
 \hline 
{} [O/Fe] & 0.29 (0.04) & 0.26 (0.04) & 0.07 (0.03) & 0.03 & 0.02 & 0.01 & flat & decrease & flat \\ [3pt]
{} [Mg/Fe] & 0.28 (0.04) & 0.25 (0.04) & 0.05 (0.03) & 0.04 & 0.03 & 0.02 & flat & decrease & flat \\ [3pt]
{} [Si/Fe] & 0.28 (0.05) & 0.18 (0.05) & 0.02 (0.03) & 0.03 & 0.02 & 0.02 & flat & decrease & flat \\ [3pt]
{} [Ca/Fe] & 0.19 (0.04) & 0.12 (0.04) & 0.05 (0.03) & 0.03 & 0.03 & 0.02 & flat & decrease & flat \\ [3pt]
{} [Cr/Fe] & 0.04 (0.06) & 0.01 (0.04) & -0.09 (0.05) & 0.05 & 0.04 & 0.03 & flat & decrease & decrease \\ [3pt]
{} [Mn/Fe] & -0.27 (0.05) & -0.12 (0.06) & 0.12 (0.08) & 0.03 & 0.03 & 0.02 & increase & increase & increase \\ [3pt]
{} [Co/Fe] & 0.04 (0.11) & 0.15 (0.10) & 0.13 (0.09) & 0.05 & 0.05 & 0.05 & increase & flat & increase \\ [3pt]
{} [Ni/Fe] & 0.08 (0.03) & 0.08 (0.03) & 0.05 (0.03) & 0.02 & 0.02 & 0.01 & flat & decrease & increase \\ [3pt]
{} [Na/Fe] & 0.03 (0.10) & -0.00 (0.11) & 0.15 (0.10) & 0.08 & 0.07 & 0.06 & decrease & flat & increase \\ [3pt]
{} [Al/Fe] & -0.00 (0.13) & 0.20 (0.08) & 0.13 (0.06) & 0.04 & 0.04 & 0.03 & increase & increase & flat \\ [3pt]
{} [K/Fe] & 0.16 (0.05) & 0.17 (0.05) & 0.06 (0.07) & 0.05 & 0.04 & 0.03 & flat & decrease & increase \\ [3pt]
\hline
\end{tabular}
\caption{
Summary statistics of Figure~\ref{fig:abundances_grid}
}
\label{tab:abundances_properties}
\tablecomments{For each element, we give the median abundance, median absolute deviation (MAD) of the abundance, median uncertainty, and qualitative behavior in each of three metallicity ranges: $\rm{[Fe/H]} < -0.8$, $-0.8 \le \rm{[Fe/H]} < 0$, and $\rm{[Fe/H]} \ge 0$.  A qualitative behavior of ``flat'' implies a linear slope with an absolute value $<$0.1.
}
\end{center}
\end{table*}

\subsubsection{Alpha Elements: O, Mg, Si, Ca}
\label{sec:alpha_elements}

{\bf Oxygen:} The inner MW [O/Fe] abundances follow the typical mean $\alpha$-element behavior: enhanced abundances, with $\rm [O/Fe] \sim +0.25$ for $\rm [Fe/H] < -0.1$, that decrease to roughly solar abundance at higher metallicities. In the range $\rm -1.0 < [Fe/H] < +0.5$, the APOGEE sample has a rather tight sequence and is less scattered than in, e.g., \citet{fulbright2006}. In both the inner Galaxy and SR samples, the oxygen abundances at a given metallicity increase with decreasing stellar temperature, a trend that is noticeable at nearly all metallicities above $\rm [Fe/H] \sim -0.9$. 
At the most metal-poor end ($\rm [Fe/H] < -1.0$), the temperature trend seems to reverse, as the O abundances become higher with {\it increasing} temperature.  

APOGEE finds lower [O/Fe] values at lower metallicity than the bulge sample of, e.g., \citet{johnson14} and other stellar samples from the literature \citep{Jonsson_2018_dr13dr14abundances}.  [O/Fe] abundances derived from observed spectra can be shifted by the inclusion of NLTE and 3D model corrections \citep[e.g.,][]{Dobrovolskas_2015_RGBmodeling}, though less work has been done to calculate the need for corrections at high metallicity.  In literature comparisons to chemical enrichment models, [O/Fe] discrepancies at low metallicity are often attributed to different models' mass cutoffs, due to the dependence of O yields on core collapse SN (CCSN) progenitor mass, or to variations in the models' assumed metallicity dependence of the CCSNe's O and Fe yields \citep[e.g.,][]{Andrews_2017_flexCE}. In APOGEE DR14/DR15, the [O/Fe] abundances at all metallicities are at roughly the same level of enhancement as [Mg/Fe] and [Si/Fe].

The nearly flat trend in [O/Fe] at increasingly higher metallicity (above ${\rm [Fe/H]=0}$) is also noteworthy; in the context of \citet{Weinberg_2017_CEevents}'s analytic chemical evolution models, this extension could indicate temporal variations in the ejected and recycled gas fractions (see, e.g., their Figure~13).  A thorough exploration of these trends in the context of different evolutionary model parameters will be explored in future work.

We also note a group of seemingly [O/Fe]-enhanced stars at high metallicities, with $\rm [O/Fe] \gtrsim +0.18$ and $\rm [Fe/H] > +0.2$ (denoted by green points in Figure~\ref{fig:abundances_grid}).  These stars tend to have lower temperatures and higher $\alpha$-element abundances than the average values of the rest of the sample, but they are not distinctly separated except in [O/Fe] and [Ca/Fe], and they show no difference in the non-$\alpha$ elements.  

After extensive investigation, we hypothesize these seemingly-enhanced abundances are most likely due to non-optimal synthetic spectral fits. The ASPCAP pipeline computes a global fit for several stellar parameters, including $T_{\rm eff}$, $\log{g}$, metallicity, and [$\alpha$/M]; these parameters are then used in the derivation of individual elemental abundances. The abundance results presented here have been derived with parameters computed using the Kurucz grid of model atmospheres \citep[the default one employed in DR14/DR15 for all of our stars;][]{Holtzman_2018_dr13dr14apogee}. In examining all of the results for these ``high-O'' stars, we observed that their global [$\alpha$/M] values are also strongly enhanced, whereas the [$\alpha$/M] values from the MARCS model grid for these stars are much more consistent with those of the rest of the sample (at a given metallicity).
The [$\alpha$/M] ratio is expected to be correlated with [O/Fe] due to the large number of OH features in cool, metal-rich stars, so this offset suggests that the O abundances for these particular stars would be better extracted using the MARCS grid.
Unfortunately, individual elemental abundances using the MARCS grid results are not available in DR14/DR15.  We exclude these stars from the analyses in the rest of the paper, but note them in the summary plots of Figure~\ref{fig:abundances_grid}, in this discussion, and as a list in Appendix~\ref{sec:app_highO} as a caveat to other users.

{\bf Magnesium:} As \citet{Schultheis_2017_apogeeBW} show for stars in Baade's Window, the APOGEE Mg abundances in the bulge region are in generally good agreement with those from {\it Gaia}-ESO and other optical studies \citep[e.g.,][]{gonzalez11a,hill2011}. \citet{Jonsson_2018_dr13dr14abundances} argues that Mg is the most precise $\alpha$-element in DR14/DR15, with practically zero offset and very small scatter compared to optically-derived values (for the same stars).

Similar to [O/Fe], we see a slight decrease in [Mg/Fe] at low metallicity (${\rm [Fe/H] \lesssim -0.8}$) for both the inner MW and SR stars. This inflection point is also seen in several simulated yields \citep[e.g.,][]{Andrews_2017_flexCE}, possibly due to the (slight) metallicity dependence of CCSNe Mg yields.
Unlike O, however, no temperature trend is visible for the Mg abundances across the entire metallicity range shown.  A small number of the ``high-O'' stars are slightly enhanced in Mg, but the majority are completely consistent with the rest of the sample.

See \S\ref{sec:knee} for an analysis of the [Mg/Fe]--[Fe/H] downturn due to contributions of Type~Ia supernovae.

{\bf Silicon:} Si is the $\alpha$-element with the smallest dispersion in our sample, especially at the metal-rich end. This low dispersion is also confirmed by other inner Galaxy studies in the infrared  \citep[e.g.,][]{rich2005,ryde2010,Schultheis_2017_apogeeBW} and in the optical \citep[e.g.,][]{fulbright2007,gonzalez11a}. However, we note a temperature trend (especially where $\rm [Fe/H] \lesssim -0.4$) in the sense that cooler stars show lower Si abundances than warmer ones at the same metallicity.  Unlike O and Mg, which plateau to their ``metal-poor'' value by $\rm [Fe/H] \sim -0.25$, [Si/Fe] continues to increase until $\rm [Fe/H] \sim -0.5$; because of the $T_{\rm eff}$ trend, [Si/Fe] of the warmer stars appears to increase more steeply at lower metallicities, but the increase is present in all temperature ranges.  Interestingly, the temperature trend appears to be much less obvious in the SR sample, but the metallicity range ${\rm [Fe/H]} <-0.5$ (where the trend is most apparent in the inner Galaxy stars) is very poorly sampled near the SR.

\begin{figure}
\begin{center}
\includegraphics[angle=0,trim=0in 0in 0in 0in, clip, width=0.45\textwidth]{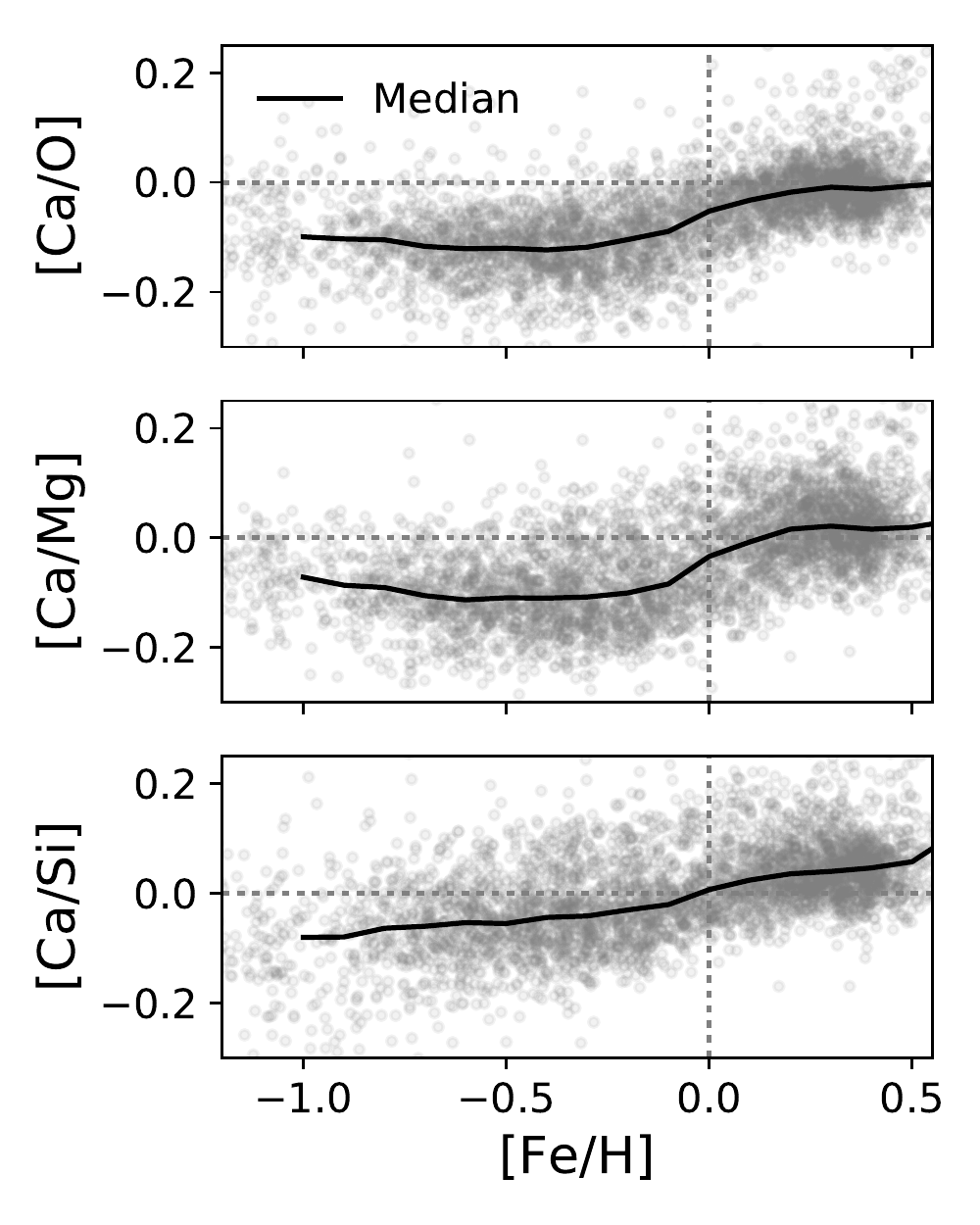} 
\caption{
Calcium abundance relative to the other $\alpha$-elements discussed here: [Ca/O] ({\it top}), [Ca/Mg] ({\it middle}), and [Ca/Si] ({\it bottom}).  The metallicity at the shift in [Ca/$\alpha$] ($\rm [Fe/H] \sim -0.2$) corresponds to the metallicity of the [Mg/Fe] ``knee'' (\S\ref{sec:knee}).
}
\label{fig:ca_versus_alpha}
\end{center}
\end{figure}

{\bf Calcium:} [Ca/Fe] exhibits slightly different behavior compared with the other $\alpha$-elements. Stars with ${\rm [Fe/H] < -0.5}$ are less enhanced in Ca (compared to solar) than in O, Mg, and Si, a phenomenon seen also in the SR sample.  Past the onset of Type~Ia SNe, which produce more Ca than other $\alpha$-elements, [Ca/Fe] declines to slightly super-solar values (more enhanced than at the SR), but [Ca/O], [Ca/Mg], and [Ca/Si] slowly rise at increasing metallicity (Figure~\ref{fig:ca_versus_alpha}), in particular at metallicities higher than the [Mg/Fe] ``knee'' discussed in \S\ref{sec:knee}. We also note a slight increase in [Ca/Fe] as [Fe/H] becomes increasingly supersolar, which could indicate a metallicity dependence in the Ca yields of CCSNe \citep[not found in subsolar metallicity progenitors;][]{Andrews_2017_flexCE} or of SN~Ias.

One striking feature is the nearly horizontal sequence of cool stars ($T_{\rm eff} < 3800$~K) with a nearly constant
[Ca/Fe] abundance of roughly $+0.25$ at high [Fe/H].  These stars overlap at the metal-richest end with the ``high-O'' stars and are most likely also related to the difficulty of analyzing certain cool stars (see the [O/Fe] discussion above).  This sequence also creates a seeming temperature trend in the metallicity range $\rm [Fe/H] > -0.5$, though the distribution becomes clearly bimodal at $\rm [Fe/H] > -0.2$ due to this likely artifact.  As for Mg, the metal-poor stars show no relation between $T_{\rm eff}$ and [Ca/Fe].

As highlighted in Table~\ref{tab:abundances_properties}, all of the $\alpha$-elements share the same qualitative behavior across metallicity: constant abundance at $\rm [Fe/H] < -0.8$, decreasing abundance from $\rm [Fe/H] > -0.8$ to $\rm [Fe/H] < 0$, and constant (near-solar) abundance for supersolar [Fe/H].

\subsubsection{Iron-peak Elements: Cr, Mn, Co, Ni}
\label{sec:fepeak_elements}

In contrast to the large body of work on the $\alpha$-elements, only a few studies exist for the abundance of Fe-peak elements in the inner MW.

{\bf Chromium:} \citet{johnson14} demonstrated that the abundance pattern of Cr in the inner MW is very similar to that of stars in the disk at larger $R_{\rm GC}$ --- i.e., roughly solar at all metallicities, though with a larger dispersion than the disk stars.  A similar conclusion was reached by \citet{Bensby_2013_bulgeages}. The APOGEE Cr abundances are the only Fe-peak abundances that have a trend indistinguishable from flat at metallicities $<$$-0.8$, with a small monotonic decrease in [Cr/Fe] up to supersolar metallicities, along with an increase in the scatter of [Cr/Fe] (Table~\ref{tab:abundances_properties}). The coloring in Figure~\ref{fig:abundances_grid} suggests that much of that apparent decrease is driven by temperature effects, with [Cr/Fe] being slightly enhanced for cooler stars with $\rm [Fe/H] < -0.5$ and slightly subsolar for cooler stars with $\rm [Fe/H] > 0.0$.
This behavior can also be seen in the coolest stars in the SR comparison sample, suggesting it may be an uncorrected trend in ASPCAP. 

Nevertheless, even considering only the warmer stars in the bulge sample (with $T_{\rm eff} > 4000$~K), a small continuous decrease of [Cr/Fe] with increasing $\rm [Fe/H]$ is detected.  In addition, [Cr/Fe] in the most metal-rich cool stars extends far lower than [Cr/Fe] in SR stars of the same temperature and metallicity.  These patterns potentially indicate a genuinely different behavior of Cr in the inner MW compared to elsewhere in the disk.
A decline in [Cr/Fe] at high metallicity is predicted by some yield models as a result of subsolar [Cr/Fe] Type~Ia SNe yields \citep[][]{Andrews_2017_flexCE}.

{\bf Manganese:} In contrast to the other elements considered in this paper, Mn has a monotonic trend of increasing $\rm [Mn/Fe]$ with increasing metallicity over all $\rm [Fe/H] > -1.5$, with [Mn/Fe] spanning $-0.27$ to $+0.12$.  The APOGEE sample is the first one in which this continuous sequence has been measured in Mn in so many stars across such a large metallicity range in the bulge.
This general pattern shape was also reported by \citet{barbuy2013} based on high-resolution optical spectra in Baade's Window for a sample of 56 stars.  They argue that the behavior of $\rm [Mn/Fe]$--$\rm[Fe/H]$ shows that Mn has not been produced under the same conditions (which may include metallicity dependence of the yields) as other iron-peak elements such as Ni. 

However, \citet{battistini15} have shown that the Mn trends can change drastically if NLTE corrections are used, resulting in  $\rm [Mn/Fe]$  becoming basically flat with metallicity.  It is not fully clear how NLTE corrections, if necessary, will affect APOGEE's giant star abundances.  The CCSN yields collated by \citet[][including \citealt{woosley1995} and \citealt{Chieffi_2004_SNyields}]{Andrews_2017_flexCE} predict a monotonic increase in [Mn/Fe], at least partially due to the increase in Mn yields at higher supernova progenitor metallicities, which is supported by the APOGEE data.  Mn production in Type~Ia SNe is also likely to be significant at higher metallicities \citep{Clayton_2003_cosmicisotopes,Andrews_2017_flexCE}.

Very striking in this distribution is the strong temperature dependency of [Mn/Fe], which causes
multiple parallel sequences separated by stellar $T_{\rm eff}$. This behavior is seen in the inner Galaxy as well as in the SR sample. What is unique about this particular temperature dependency, compared to others in this paper, is the parallel nature of the sequences, which strongly suggests that the {\it shape} of the trend is robust and that the broad span is due to vertical $T_{\rm eff}$-dependent offsets.  These offsets may be related, in part, to the large temperature trend in pre-calibrated Mn abundances described in \citet{Holtzman_2018_dr13dr14apogee} and/or to the temperature-dependent NLTE corrections discussed in \citet{Bergemann_2008_MnNLTEeffects}.

We note that the sharp angled edge to the stellar distribution at $\rm [Fe/H] \sim +0.5$ is due to the edge of the range over which the Mn calibrations are valid \citep{Holtzman_2018_dr13dr14apogee}.

{\bf Cobalt:} The DR14/DR15 Co abundances have a wavy ``cubic'' pattern and are in general enhanced relative to solar over the entire metallicity range, with significant dispersion (and correspondingly higher uncertainties). The median [Co/Fe] peaks at $+0.2$ near $\rm [Fe/H] \sim -0.5$, drops to about $+0.05$ at solar metallicity (producing the ``flat'' trend in the middle metallicity bin of Table~\ref{tab:abundances_properties}), and then increases back to $+0.3$ before the calibration cutoff just below $\rm [Fe/H] = +0.5$. A similar pattern is seen for the SR sample, particularly at higher [Co/Fe], albeit with a slightly smaller dispersion. Stars of all temperatures show this same pattern shape and high dispersion, with cooler stars at lower metallicities ($\rm -1.0 < [Fe/H] < -0.2$) being slightly Co-enhanced relative to their warmer counterparts. We note that \citet{Holtzman_2018_dr13dr14apogee} also describe $T_{\rm eff}$ trends with [Co/Fe] in clusters and advocate caution when using Co abundances, and \citet{Jonsson_2018_dr13dr14abundances} discuss potential metallicity-dependent offsets from literature values.

To our knowledge, only one similar study of this element in the inner Galaxy exists; \citet{johnson14} find behavior qualitatively similar in shape but less clearly-defined, possibly due to their smaller sample size. 
We note that this pattern is not generally observed in samples of solar neighborhood disk stars \citep[e.g.,][]{battistini15}, which, considered in combination with the caveats stated above, support caution when interpreting DR14/15 Co abundances.

Together with Mn, Co is modeled as being produced by explosive silicon burning in CCSNe \citep{woosley1995} and to a smaller extent in SN~Ias \citep{bravo12}.  Taking the DR14/15 abundances at face value, the rise in [Co/Fe] with [Fe/H] at lower metallicities is consistent with [Fe/H]-dependent CCSN yields.  The decrease in [Co/Fe] between $\rm [Fe/H] = -0.5$ and solar is due to the contribution from SNe~Ia, which have a lower [Co/Fe] ratio than CCSNe in this metallicity range, much in the same manner as the $\alpha$-elements \citep{Andrews_2017_flexCE}.  The upturn in [Co/Fe] for ${\rm [Fe/H]>0}$ supports a continued [Fe/H] dependence in CCSN yields at high metallicities.

{\bf Nickel:} [Ni/Fe] has a morphology qualitatively similar to [K/Fe] but with a much smaller amplitude variation and smaller dispersion, more akin to [Mg/Fe].  \citet{Jonsson_2018_dr13dr14abundances} finds that Ni abundances are the most precise iron-peak abundances in APOGEE, based on comparison to literature studies stars in common with APOGEE.  

The mean [Ni/Fe] increases slightly from low metallicities to peak near $+0.1$ at $\rm [Fe/H] \sim -0.7$, like K and Mg, before dropping to a minimum at solar metallicity and then rising again.  The cooler stars extend to higher [Ni/Fe] values at higher metallicities, but beyond that, we do not find any significant differences between either the inner Galaxy and SR samples or between stars with different temperatures. A similar behavior in Ni was found by \citet{johnson14}, including the upturn to higher Ni as [Fe/H] approaches $+0.5$.  Simulations that include metallicity-dependent CCSN yields predict a monotonic increase in [Ni/Fe] at low metallicity, which is not strongly supported by our data, though these could be responsible for the small increase between $-1.5 \le {\rm [Fe/H]} \le -0.5$. The upturn in [Ni/Fe] at supersolar metallicities could be due to a metallicity dependence of SN~Ia [Ni/Fe] yields, on top of the CCSN contributions.

\subsubsection{Odd-Z Elements: Na, Al, K}
\label{sec:oddz_elements}

{\bf Sodium:} The DR14/DR15 [Na/Fe] abundances exhibit a large scatter, larger than typical uncertainty.  This is due (at least in part) to the presence of strong telluric absorption near one or both of the abundance windows in many stars; the impact of this absorption depends on the radial velocity of each star and is not included in the uncertainty calculation described in \S\ref{sec:sample} and \citet{Holtzman_2018_dr13dr14apogee}.
As in [Mn/Fe], there appears to be a monotonic increase in mean [Na/Fe] with increasing metallicity, though the median trend (Figure~\ref{fig:abundances_grid_sn}) cannot be distinguished from flat at ${\rm [Fe/H] \lesssim -0.2}$. A similar trend is seen by \citet{Bensby_2017_bulgedwarfsSGs} in a sample of microlensed dwarf stars in the bulge. However, as \citet{smiljanic16} show, none of the chemical evolution models can explain the observed increase of $\rm [Na/Fe]$ for $\rm [Fe/H] > 0$, which may suggest that the models lack some site of Na production at later stages or that the metallicity dependence of CCSN Na yields is underestimated.

\citet{smiljanic16} discuss the strong NLTE effects that Na lines can display, leading to corrections of $\lesssim$0.2~dex. 
However, \citet{Cunha_2015_NaOinngc6791} find very small NLTE corrections for the $H$-band lines used in APOGEE.  We do not observe any temperature trend in the derived Na abundances, but a larger intrinsic scatter of [Na/Fe] is apparent, compared to the solar radius sample.

As in the [Mn/Fe] distribution (\S\ref{sec:fepeak_elements}), the sharply angled cutoffs to the [Na/Fe] distribution at $\rm [Fe/H] \sim +0.5$ and $\rm [Fe/H] \sim -1$ are due to the edge of the range over which the Na calibrations are valid \citep{Holtzman_2018_dr13dr14apogee}.

{\bf Aluminum:} The Al abundance distributions appear at first glance to differ greatly between the inner MW and SR samples. In the inner MW sample, the most metal-poor stars have subsolar [Al/Fe] values, increasing monotonically to $\rm [Al/Fe] \sim +0.2$ at $\rm [Fe/H] \sim -0.25$, and then indistinguishable from a flat trend at higher metallicities.  In contrast, the SR sample has two roughly parallel sequences, both with [Al/Fe] increasing monotonically at higher metallicities but offset from each other at a given [Fe/H] by $\sim$0.2~dex in [Al/Fe].  However, we believe these seemingly different patterns are driven simply by the lack of metal-poor, $\alpha$-poor stars in the inner Galaxy; the densely-populated $\alpha$-rich inner MW stars follow the same behavior as the $\alpha$-rich SR stars, and the metal-rich $\alpha$-poor stars in both groups occupy the same [Al/Fe]--[Fe/H] space.  

A slight temperature trend is visible for stars with $\rm [Fe/H] \lesssim -0.4$, in the sense that warmer stars have higher [Al/Fe] values at a given [Fe/H], and the ``calibration-range'' edge is visible at $\rm [Fe/H] \sim +0.5$ (as in Na, K, Mn, and Co).
The global scatter is slightly larger than would be expected based on the typical uncertainty, which is due in part to the presence of two parallel sequences, and in part to the sensitivity of the Al lines to $T_{\rm eff}$ and NLTE effects \citep[e.g.,][]{Hawkins_2016_APOGEEabundances,Nordlander_2007_AlNLTEeffects,Jonsson_2018_dr13dr14abundances}.

While studies generally agree that [Al/Fe] is enhanced for stars with $\rm [Fe/H] < -0.3$ \citep[e.g.,][]{johnson14}, there is less consensus regarding Al abundances at supersolar metallicities. For example, \citet{fulbright2007}, \citet{lecureur07}, and \citet{Alves-Brito_2010_bulgethickdisk} find enhanced [Al/Fe] at high metallicity, while \citet{johnson12} and \citet{Bensby_2013_bulgeages} observe a decline in [Al/Fe] similar to that seen in the $\alpha$-element abundances. \citet{johnson14} find Al to behave similar to the $\alpha$-elements Mg, Si, and Ca, while we see in the APOGEE data a clear difference between the abundances of Al and the $\alpha$-elements at subsolar metallicities. At supersolar metallicities, the dispersion is too large to identify any trend.  

Many stellar yield models assume that Al production is similar to that of Na. However, these yields have been shown to poorly represent the observed behavior of [Al/Fe] in the disk (see Figure~9 in \citealt{Andrews_2017_flexCE}), which in turn is used as evidence that the metallicity dependence of Al production is weaker than in those yield models. The APOGEE data could be interpreted as arguing against this conclusion, because the observed [Al/Fe]--[Fe/H] distribution is qualitatively well-matched by the more strongly metallicity-dependent models of \citet{Chieffi_2004_SNyields} as presented in \citet{Andrews_2017_flexCE}, but the large scatter renders this a relatively weak argument.

{\bf Potassium:} At subsolar metallicities, K has typical $\alpha$-element behavior, with an almost-flat plateau at metallicities less than $-0.8$ (median $\rm [K/Fe] = +0.16$), followed by a decrease in [K/Fe] {\bf to $+0.06$} with increasing metallicity (especially in the warmer stars). For the most metal-rich stars (with $\rm [Fe/H] > +0.2$), [K/Fe] increases again.  The prominence of this increase is most likely artificial, originating from the contribution of the coolest and most metal-rich stars ($T_{\rm eff} < 3800$~K, yellow in Figure~\ref{fig:abundances_grid}).  However, even the warmer stars show a small ($\sim$0.1 dex) increase at $\rm [Fe/H] > +0.3$, seen in both the inner MW and SR samples.  As in the $\alpha$-elements and Al, the most significant differences between the two samples arises in the range $\rm -0.5 < [Fe/H] < 0.0$, where the low-$\alpha$ (and low-[K/Fe]) SR stars have no counterpart in the inner MW (see also \S\ref{sec:sn_comparison}).  In this way, [K/Fe] again shows similarity to the $\alpha$-elements.

Chemical yield models assume a relatively weak but non-zero metallicity dependence for [K/Fe] production in CCSNe. The similarity of the [K/Fe]--[Fe/H] distribution to that of [Mg/Fe] and other $\alpha$-elements with a slight metallicity dependence in their yields supports this assumption, though the overall K yields are in general underpredicted by CCSN yield models \citep{Andrews_2017_flexCE}.  The increase in [K/Fe] at high [Fe/H], similar to that in [Ni/Fe], is not predicted by the models and may be evidence for a stronger CCSN yield metallicity dependence than assumed and/or a non-negligible contribution from SN~Ias.

\subsection{Comparison to the Solar Radius Sample}
\label{sec:sn_comparison}

\begin{figure}[!hptb]
\begin{center}
\includegraphics[angle=0,trim=0in 0in 0in 0in, clip, width=0.45\textwidth]{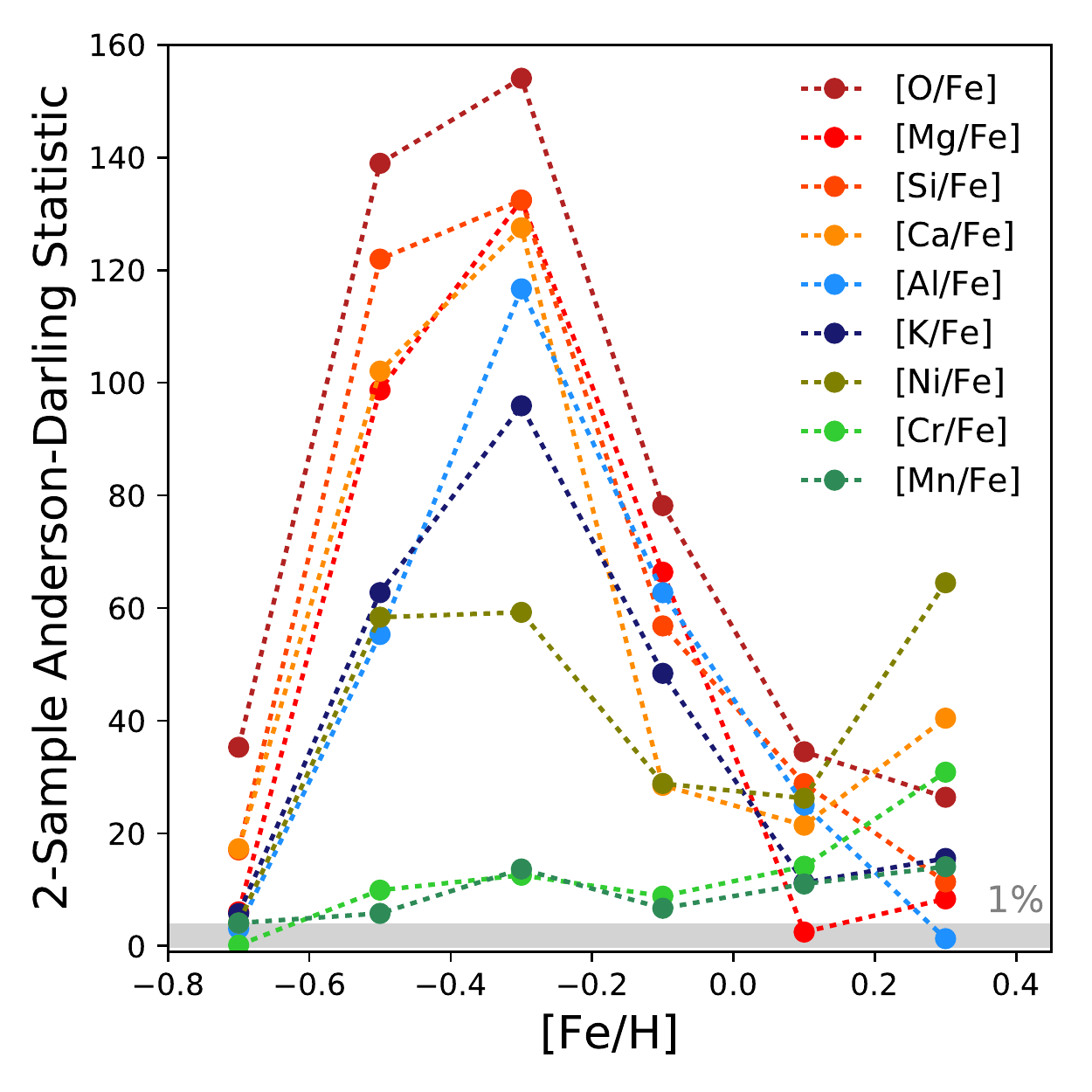} 
\caption{
The two-sample Anderson-Darling statistic for the inner Galaxy and SR abundances as a function of [Fe/H].  The $\alpha$-elements (red/orange shades) show the greatest differences, particularly in the metallicity range where the SR disk stars have a bimodal distribution of $\alpha$-element abundances.  Odd-Z elements have blueish colors, and Fe-peak elements are shaded green.
The gray band shows the critical value above which the null hypothesis can be rejected at the 1\% significance level.
}
\label{fig:cp_SN_stats}
\end{center}
\end{figure}

Figure~\ref{fig:cp_SN_stats} shows the two-sample Anderson-Darling statistic \citep{Scholz_1987_ksampleandersondarling} for the abundance distributions [X/Fe] of the inner MW and SR samples, measured at different [Fe/H]. This statistic measures the likelihood that given samples of data are drawn from the same parent distribution. Because of the difficulties of matching the inner MW's coolest, most metal-poor stars at the SR (Appendix~\ref{sec:app_snsample}), we restrict these compared samples to $T_{\rm eff} \ge 3800$~K and $\rm [Fe/H] \geq -1.0$, and remove bins without at least 30 stars in both samples.  We also do not show Na and Co, the elements with the largest dispersions. 
The exact values of the statistic on the y-axis are less informative than the relative trends across metallicity and between elements.  The horizontal gray bar at the bottom of Figure~\ref{fig:cp_SN_stats} indicates the critical value above which the hypothesis that these samples are drawn from the same distribution can be rejected at the 1\% significance level.

As expected, the $\alpha$-elements are the most dissimilar in the metallicity range $\rm -0.5 \lesssim [Fe/H] \lesssim -0.1$, where the SR stars have a bimodal distribution in [$\alpha$/Fe] and the inner MW stars have a single sequence.  In this same range, Al, K, and Ni also exhibit differences, smaller than those in O, Mg, Si, and Ca.  Both the $\alpha$-elements and these three are more similar between the two samples at lower and higher metallicities, though [Ca/Fe] and [Ni/Fe] show increased differences at the highest metallicities.  In the case of [Ca/Fe], this may be due to the increased dispersion in the inner MW stars compared to the SR and to the smattering of cooler stars (near the 3800~K cutoff) with seemingly enhanced [Ca/Fe].  The differences in [Ni/Fe] at high [Fe/H] are most likely due to the steeper upturn in [Ni/Fe] in the inner Galaxy stars, again potentially driven by the greater fraction of cooler stars. In contrast, the distributions of the Fe-peak elements Cr and Mn are measured to be much more similar between the inner MW and SR samples at all metallicities.

\begin{figure*}[!hptb]
\begin{center}
\includegraphics[angle=0,trim=0in 0in 0in 0in, clip, width=0.9\textwidth]{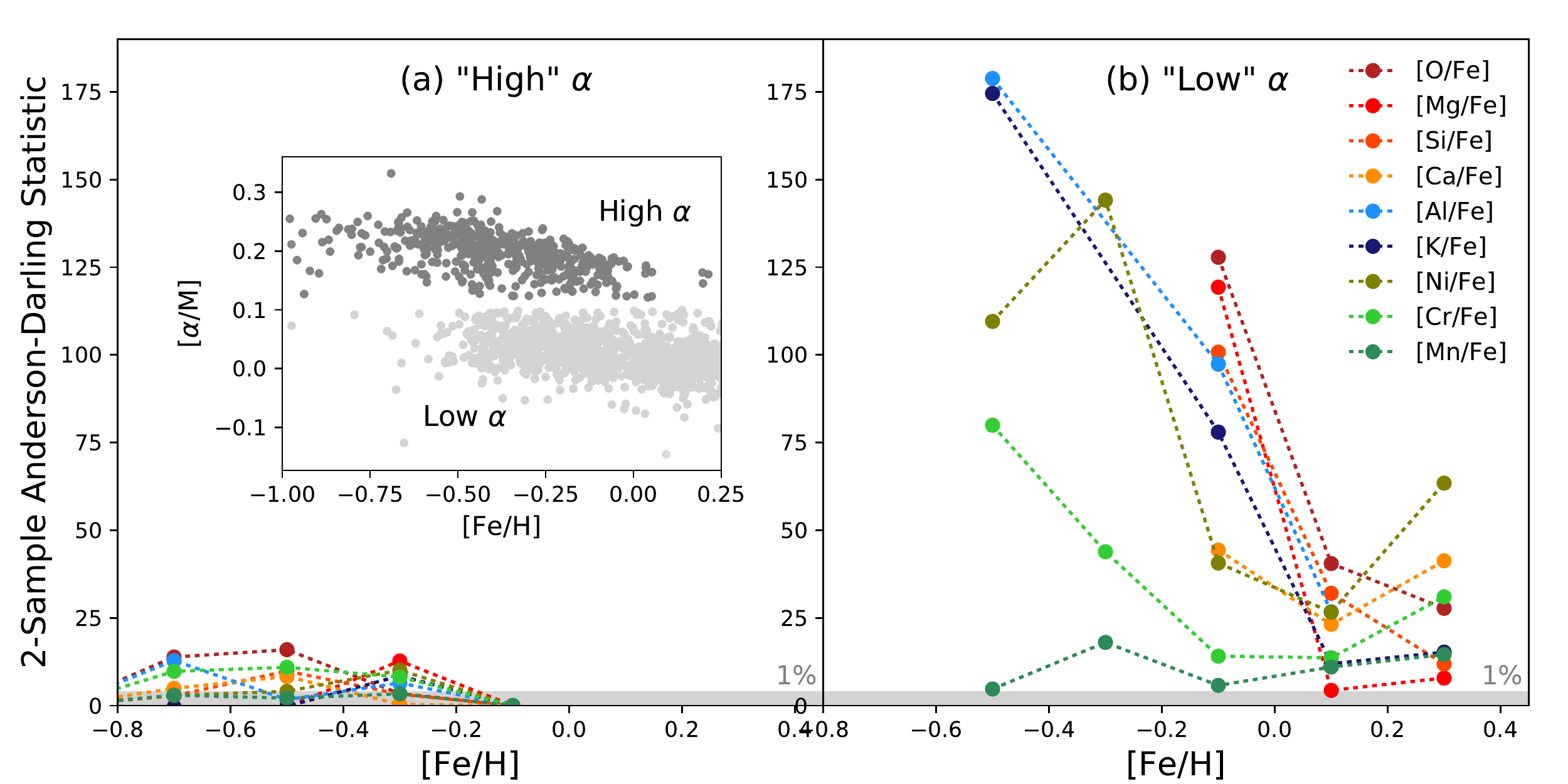} 
\caption{
As in Figure~\ref{fig:cp_SN_stats}, the 2-sample Anderson-Darling statistic for the inner Galaxy and SR abundances as a function of [Fe/H], separately for the high-$\alpha$ abundance stars ({\it left}; ${\rm [\alpha/M]} > +0.12$) and low-$\alpha$ abundance stars ({\it right}; ${\rm [\alpha/M]} < +0.1$).  The color of each element is identical to that in Figure~\ref{fig:cp_SN_stats}, {\bf as is the 1\% significance band at the bottom}.  Not all elements appear in both panels due to the small number of stars at certain [$\alpha$/M]--[Fe/H] coordinates.  The inset in the left panel shows the [$\alpha$/M]--[Fe/H] distribution for the high-$\alpha$ abundance (dark gray) and low-$\alpha$ abundance (light gray) stars.
}
\label{fig:cp_SN_stats_highlowalpha}
\end{center}
\end{figure*}

In Figure~\ref{fig:cp_SN_stats_highlowalpha}, this statistic is computed separately for the SR's ``high-$\alpha$'' and ``low-$\alpha$'' sequences (shown in dark and light gray, respectively, in the left panel's inset), in comparison with the inner MW.  We define ``high-$\alpha$'' here as $\rm [\alpha/M] > +0.12$ and ``low-$\alpha$'' as $\rm [\alpha/M] < +0.1$, to emphasize the differences.  The trend lines are noisier due to the smaller sample sizes, and the restricted metallicity ranges means measurements cannot be made in all [Fe/H] bins.  Nevertheless, it is clear that the SR's ``high-$\alpha$'' stellar abundances are more similar to the inner MW, even for non-$\alpha$ elements. Similarly, the ``low-$\alpha$'' SR stars have elemental abundances that are, in general, more discrepant from the inner MW's stars at the same metallicity.  (Cr again appears to be the exception.)  A more thorough investigation of the relationship between chemistry and kinematical properties in both samples will shed light on the relationship between the inner bar/bulge structure and the inner Galactic disk.

\subsection{Location of the [Fe/H] ``Knee''}
\label{sec:knee}

The shape of a stellar population in the canonical [$\alpha$/Fe]--[Fe/H] plane contains information about the efficiency and duration of star formation.  In particular, when the Type~Ia~SNe begin to explode and contribute their higher yields of Fe (relative to the $\alpha$ elements), the [$\alpha$/Fe] ratios of new stars drop, even as [Fe/H] continues to increase.  
The downturn imprinted in the sequence by this [$\alpha$/Fe] decrease is often referenced colloquially as the ``knee''.  The [Fe/H] value at which this occurs marks the point at which SNe~Ia events become a significant source of Fe, which in turn depends on the early star-formation rate of the population \citep[e.g.,][]{Matteucci_2003_chemicalevolution,Matteucci_2012_chemicalevolution}.

\citet{johnson14} compared the knee position of the Galactic bulge to that of the local thick disk and found that the bulge knee position lies at a higher [Fe/H] value compared with the thick disk, which is in agreement with measurements by \citet{Bensby_2017_bulgedwarfsSGs} based on bulge dwarf stars. However, \citet{Bensby_2017_bulgedwarfsSGs} also points out there are inconsistencies in the direction of the knee position's metallicity shift between Mg \& Ca (positive shifts with $R_{\rm GC}$) and Si \& Ti (negative shifts).
In \citet{Nidever_2014_MWchemevolution} and \citet{hayden15}, the near-constancy of the knee's position across a large range of $R_{\rm GC}$ (3--15~kpc) is interpreted as qualitative evidence for spatial and chemical homogeneity of the star-forming disk at early times.  In \citet{RojasArriagada_2017_GESbulge}, the position of the $\alpha$-enhanced disk's [Mg/Fe]--[Fe/H] downturn was measured quantitatively over a range of $R_{\rm GC}$ using {\it Gaia}-ESO stars, and was also found to be consistent with a constant [Fe/H] position (though with a potentially significant shift to lower [Fe/H] at radii beyond $R_{\rm GC} \approx 8$~kpc).  In that study, all stars with $R_{\rm GC} < 4$~kpc were considered in a single bin.  Here, we use our sample of $R_{\rm GC} < 4$~kpc stars to measure the position of the [Mg/Fe]--[Fe/H] downturn\footnote{[Mg/Fe] was chosen for this measurement because the [Mg/Fe]--[Fe/H] trend has the smallest dispersion and the clearest turnover of all of the $\alpha$-elements (Figure~\ref{fig:abundances_grid}).  Accordingly, [Mg/Fe] is also deemed the most accurate DR14/DR15 $\alpha$-element abundance by \citet{Jonsson_2018_dr13dr14abundances}.} in multiple radial bins (with $\Delta R_{\rm GC} = 0.75$~kpc).  

\begin{figure*}[!hptb]
\begin{center}
\includegraphics[angle=0,trim=0in 0in 0in 0in, clip, width=\textwidth]{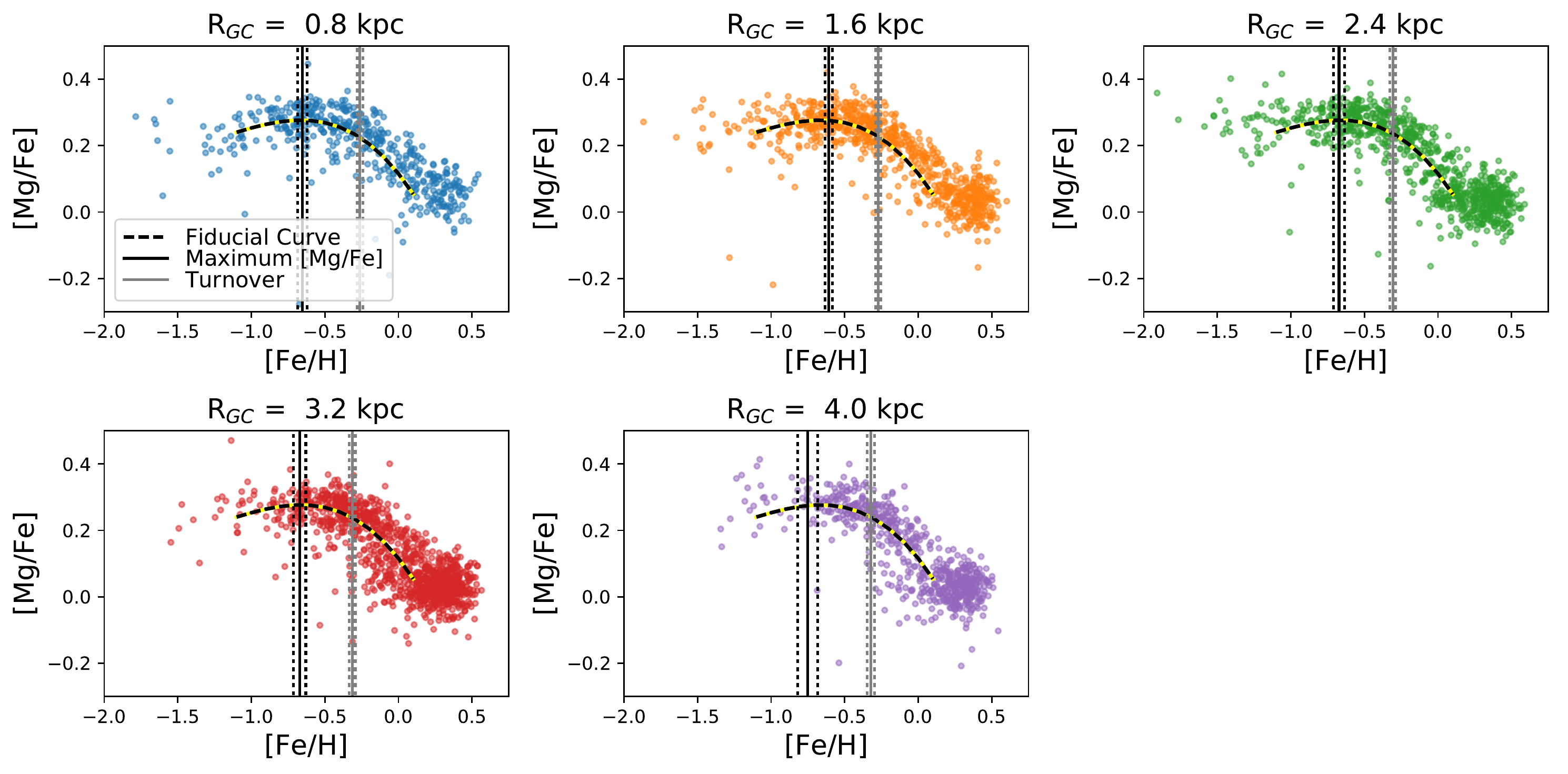} 
\caption{
Measurement of [Fe/H] at the [Mg/Fe] downturn --- i.e., the metallicity of the $\alpha$-element knee attributable to the contributions of Type~Ia~SNe.  The points are the stars in each $R_{\rm GC}$ bin as labeled.  
The black and gray vertical lines indicate two different indicators of this downturn, as described in the text.  The black/yellow dashed line is a fiducial cubic fit to the full [Mg/Fe]--[Fe/H] distribution, identical in all panels.
}
\label{fig:knee_position_panels}
\end{center}
\end{figure*}

\begin{figure}
\begin{center}
\includegraphics[angle=0,trim=0in 0in 0in 0in, clip, width=0.5\textwidth]{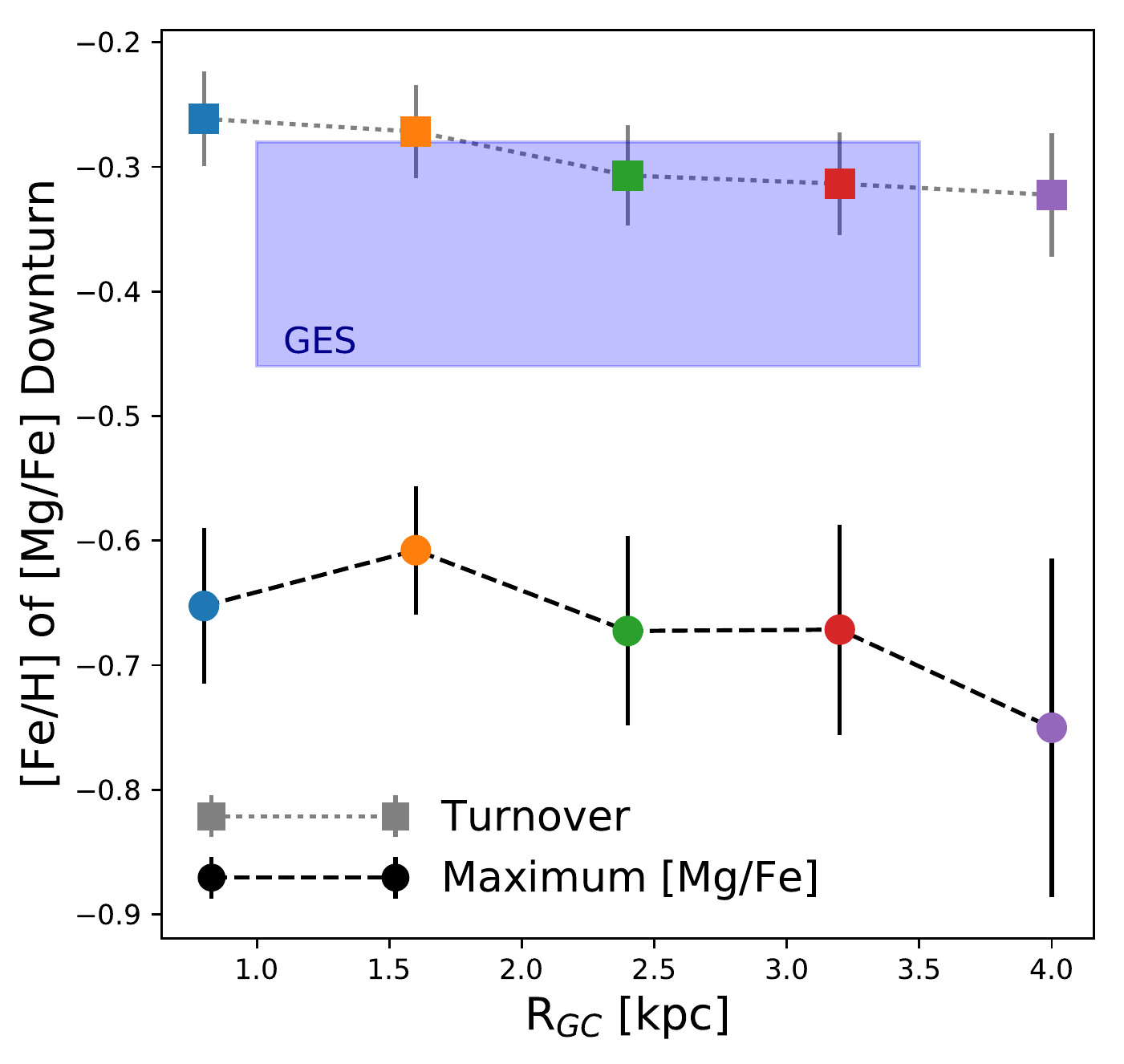} 
\caption{
Trend of the [Mg/Fe] downturn positions highlighted in Figure~\ref{fig:knee_position_panels} as a function of $R_{\rm GC}$.  The square points correspond to the ``turnover'' metric shown in gray in Figure~\ref{fig:knee_position_panels} and described in the text, and the circles correspond to the peak [Mg/Fe] metallicity shown in black in Figure~\ref{fig:knee_position_panels}.  The blue box labeled ``GES'' corresponds to the bin size and downturn metallicity range measured in the Gaia-ESO sample by \citet{RojasArriagada_2017_GESbulge}.
}
\label{fig:knee_position}
\end{center}
\end{figure}

There exist multiple metrics by which one could define the location of this turnover.  \citet{RojasArriagada_2017_GESbulge} use the intersection of two straight lines, one fit to the high-[Mg/Fe] sequence and one to the low one, but for the APOGEE dataset's distribution, we found this method to be rather sensitive to outliers and to the choice of where to separate high- and low-[Fe/H] stars.  We explored a wide range of alternative ``[Mg/Fe] downturn'' indicators, two of which are demonstrated in Figures~\ref{fig:knee_position_panels}--\ref{fig:knee_position}.  

Figure~\ref{fig:knee_position_panels} contains the [Mg/Fe]--[Fe/H] distributions of the stars in different $R_{\rm GC}$ bins, as labeled, overplotted with the same fiducial cubic polynomial in each bin (black--yellow dashed line).  The two [Mg/Fe] downturn metrics demonstrated here are calculated using $N=500$ independent realizations in which the $R_{\rm GC}$, [Fe/H], and [Mg/Fe] values of each star are drawn from normal distributions centered on the star's nominal values with standard deviations equivalent to the star's $R_{\rm GC}$, [Fe/H], and [Mg/Fe] uncertainties.  Thus the exact set of stars in each $R_{\rm GC}$ bin, and their [Mg/Fe]--[Fe/H] distribution, are allowed to vary between realizations.  The median of the 500 measurements is taken as the final value of each metric, with an uncertainty of 3$\times$ the median absolute deviation (MAD) of the 500 measurements.

The black vertical line in each panel of Figure~\ref{fig:knee_position_panels} indicates the local maximum of a cubic polynomial, as determined from the first derivative, fitted to just the stars in that $R_{\rm GC}$ bin.  We restricted the fitting range to $-1.0 \le {\rm [Fe/H]} \le +0.1$, to avoid variations in the fits driven by the different numbers of stars with ${\rm [Fe/H]} < -1.0$ in the bins. These values are plotted against $R_{\rm GC}$ in Figure~\ref{fig:knee_position} with colored circles connected with a dashed black line.  The accompanying vertical black dashed lines in Figure~\ref{fig:knee_position_panels}, shown as uncertainty bars in Figure~\ref{fig:knee_position}, are the 3$\times$MAD values from the $N=500$ realizations.  

The gray vertical lines in Figure~\ref{fig:knee_position_panels} labeled ``Turnover'' correspond to the [Fe/H] value at which the local derivative, $\sfrac{d{\rm [Mg/Fe]}}{d{\rm [Fe/H]}}$, equals $-0.25$.  This derivative value, though semi-arbitrary, was chosen to reproduce where a set of human observers visually identified the downtown in the full sample's [Mg/Fe]--[Fe/H] distribution, but, unlike visual identification, it can be computed quantitatively for any subset of stars.  These [Fe/H] values are plotted in Figure~\ref{fig:knee_position} as colored squares connected by a gray dotted line; as with the local maximum metric, the uncertainty bars on the turnover metric are the 3$\times$MAD values resulting from the multiple data realizations.

Both of these metrics produce [Mg/Fe] downturn positions that are largely constant with $R_{\rm GC}$, with a potential decrease to lower metallicities at larger radii.  We do not consider this decrease strongly significant, since the ``turnover'' values are mutually consistent within the uncertainties of the inner and outermost bins, and the most metal-poor ``maximum [Mg/Fe]'' points have noticeably larger uncertainties (due to the relative dearth of metal-poor stars in those bins to firmly anchor the polynomial).  We note that the decrease observed is within the uncertainties of the downturn position measured by \citet{RojasArriagada_2017_GESbulge} in their single bulge bin, shown as the blue box in Figure~\ref{fig:knee_position}.
We have also confirmed that the placement and size of the bins themselves have no impact on the conclusions drawn.

Other downturn indicators that we evaluated include the cubic fit's inflection point (i.e., $\sfrac{d^2{\rm [Mg/Fe]}}{d{\rm [Fe/H]}^2}=0$), the metallicity at which the cumulative distribution function of metallicities of the high-[Mg/Fe] stars reaches some selected value, and alternative values of $\sfrac{d{\rm [Mg/Fe]}}{d{\rm [Fe/H]}}$.  All of these produce downturn positions that are reasonable approximations to what previous efforts have historically termed the ``knee'', and which are either flat with $R_{\rm GC}$ or exhibit a small decrease as in Figure~\ref{fig:knee_position}.  There is no widely accepted definition of this morphological feature against which to test these different metrics, so we emphasize here the mean behavior and defer a detailed assessment of the metrics in the context of physical interpretation to future work.

The approximate constancy of the downturn positions is indicative of either similar star-formation environments across a large fraction of the young Galaxy, perhaps due to a well-mixed ISM in the early star-forming disk and bulge region \citep[e.g.,][]{Bournaud_2009_highzthickdisks}, or of significant radial mixing since that era \citep[e.g.,][]{Minchev_2013_diskCDevolution} that has smoothed out any initial trends due to gradients in the star-formation rate.  The former is the conclusion reached by \citet{Nidever_2014_MWchemevolution} and \citet{hayden15} for the high-$\alpha$ sequence at larger $R_{\rm GC}$ beyond the bulge, but the potential importance of radial mixing in the densely packed inner MW cannot be discounted \citep[e.g.,][]{Loebman_2016_MDFradialmigration}. Indeed, if confirmed, the slight shift of the downturn position to lower metallicities at larger $R_{\rm GC}$ would support the latter scenario operating at some level, since a well-mixed ISM with no radial migration would not easily produce a coherent gradient. Such a downturn could also result from a gradient in star formation rate, even in a chemically well-mixed ISM.
Recently, \citet{Mackereth_2018_simalphaelements} showed that the properties of the high-$\alpha$ sequence are relatively consistent throughout the EAGLE simulations, even as the presence of a bimodal sequence (as seen at larger $R_{\rm GC}$ in the MW, including in our SR sample) are rare in the simulations.

\subsection{[X/Fe] Correlations}
\label{sec:correlations}

The large sample size, in terms of number of elements and number of stars, gives us increased statistical power for constraining the correlation between pairs of elements in different populations of stars.  These are useful measures for identifying similarities and differences driven by nucleosynthetic yields from various pathways (including the metallicity dependence of CCSN and SN~Ia yields), and they also provide a simple way to parameterize the high-dimensional chemical space for data-model comparison. In this section, we explore the linear correlation between pairs of elements for the entire sample and for high- and low-[$\alpha$/Fe] stars separately.  In the latter case, we focus on pairs of elements whose correlation exhibits interesting similarities or significant differences between the two groups.

\begin{figure*}[!hptb]
\begin{center}
\includegraphics[angle=0,trim=0in 0in 0in 0in, clip, width=0.9\textwidth]{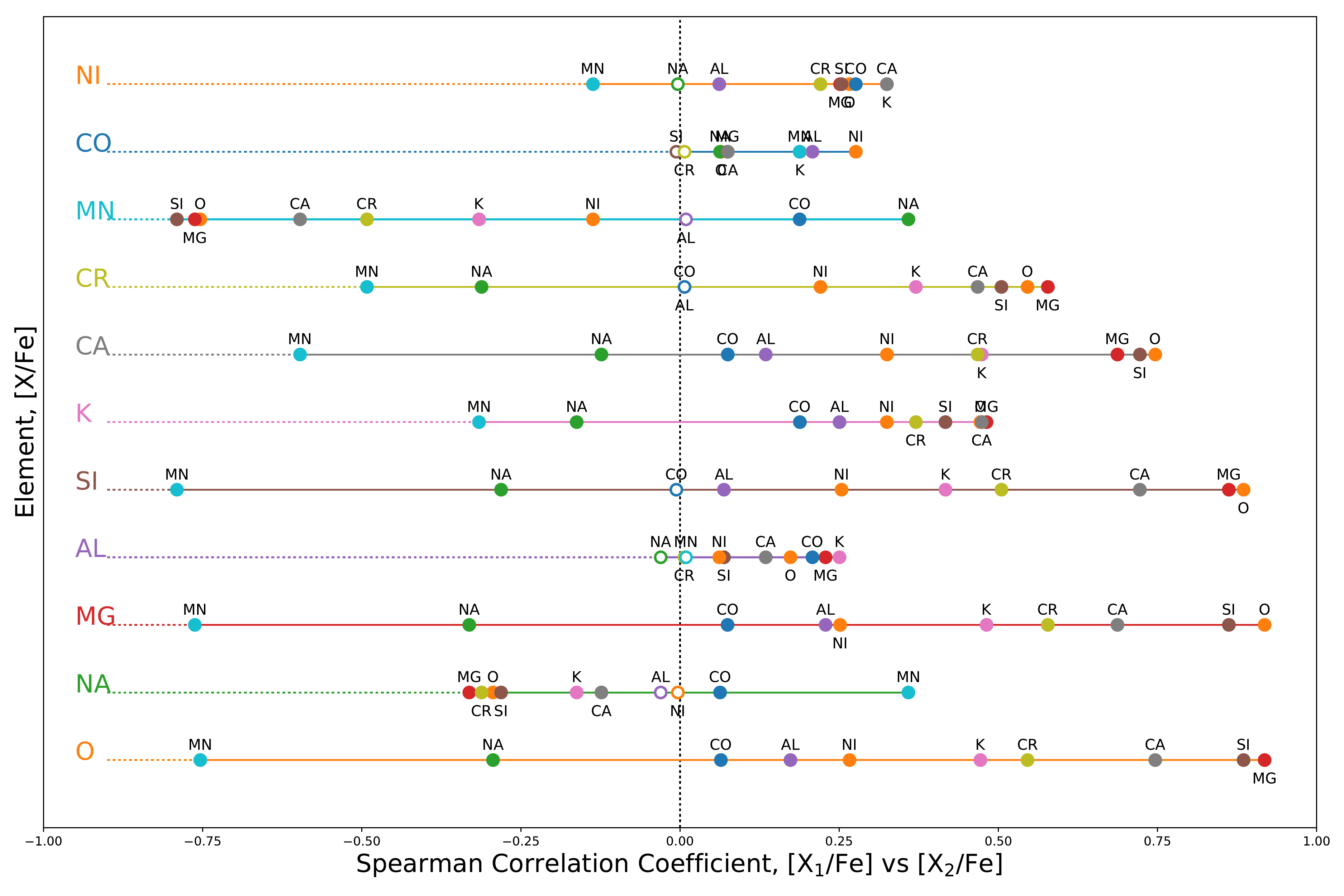} 
\caption{
Spearman correlation coefficient, $\rho_S$, for all of the stars in our inner MW sample. Values of $\rho_S$ near $+1.0$ and $-1.0$ indicate strong positive and negative linear correlation, respectively.  Pairs of distributions with $\rho_S \sim 0$ cannot be distinguished from an uncorrelated dataset, and empty circles indicate low-significance correlations with a $p$-value of $\geq$0.05.
}
\label{fig:element_correlations}
\end{center}
\end{figure*}

Figure~\ref{fig:element_correlations} is a visualization of the Spearman linear correlation coefficient \citep[$\rho_S$; e.g.,][]{Kokoska_2000_ProbStats} for all pairs of elements.  Each horizontal line represents one element, and points are placed along it at the $\rho_S$ correlation value between that element and the one labeled above the point. The color of each point is the same as that element's horizontal line. For example, the bottom line represents [O/Fe], and the points along that line indicate the linear correlation coefficient between [O/Fe] and [X/Fe] for each of the other elements.  This line for [O/Fe] is orange, so the points indicating [O/Fe]'s correlation with the other elements are also orange.  Points for correlations with a $p$-value of $\geq$0.05 are shown as empty circles.

The $\alpha$-elements are well-correlated with each other, as expected; Ni, K, and Cr are also correlated with the $\alpha$-elements, though more weakly.  One dramatic feature is the anti-correlation of [Mn/Fe] with most other elements, especially the $\alpha$-elements, but including the other iron-peak elements Cr and Ni. This trend should be even stronger if the $T_{\rm eff}$ dependency seen in [Mn/Fe] (\S\ref{sec:fepeak_elements}) were taken into account.  On the other hand, Mn shows a positive correlation with Co; the abundances of both elements increase with increasing [Fe/H] at low ($<-0.5$) and high ($>0.0$) metallicities, likely due to metallicity-dependent CCSN yields.

\begin{figure*}[!hptb]
\begin{center}
\includegraphics[angle=0,trim=0in 0in 0in 0in, clip, width=0.4\textwidth]{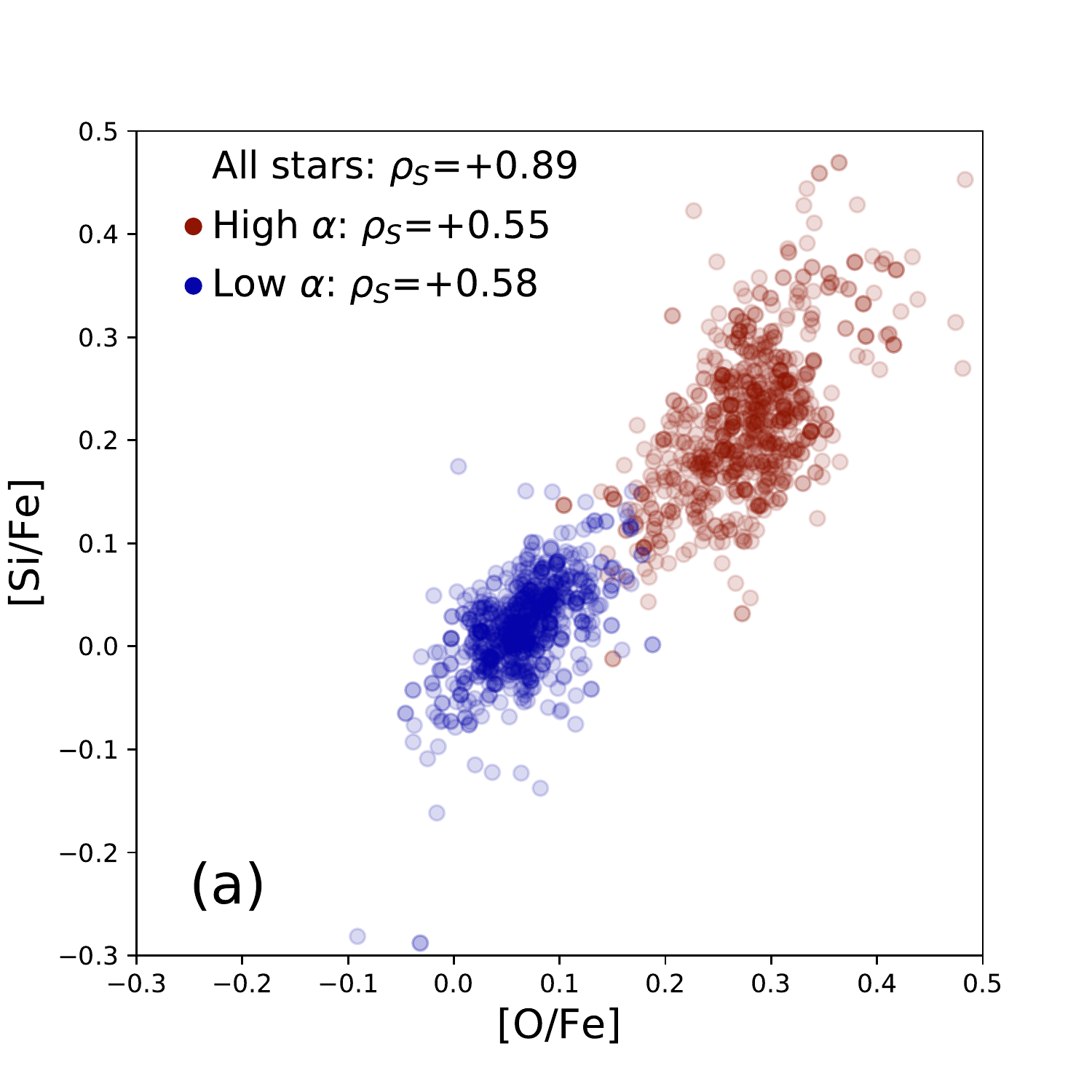} 
\includegraphics[angle=0,trim=0in 0in 0in 0in, clip, width=0.4\textwidth]{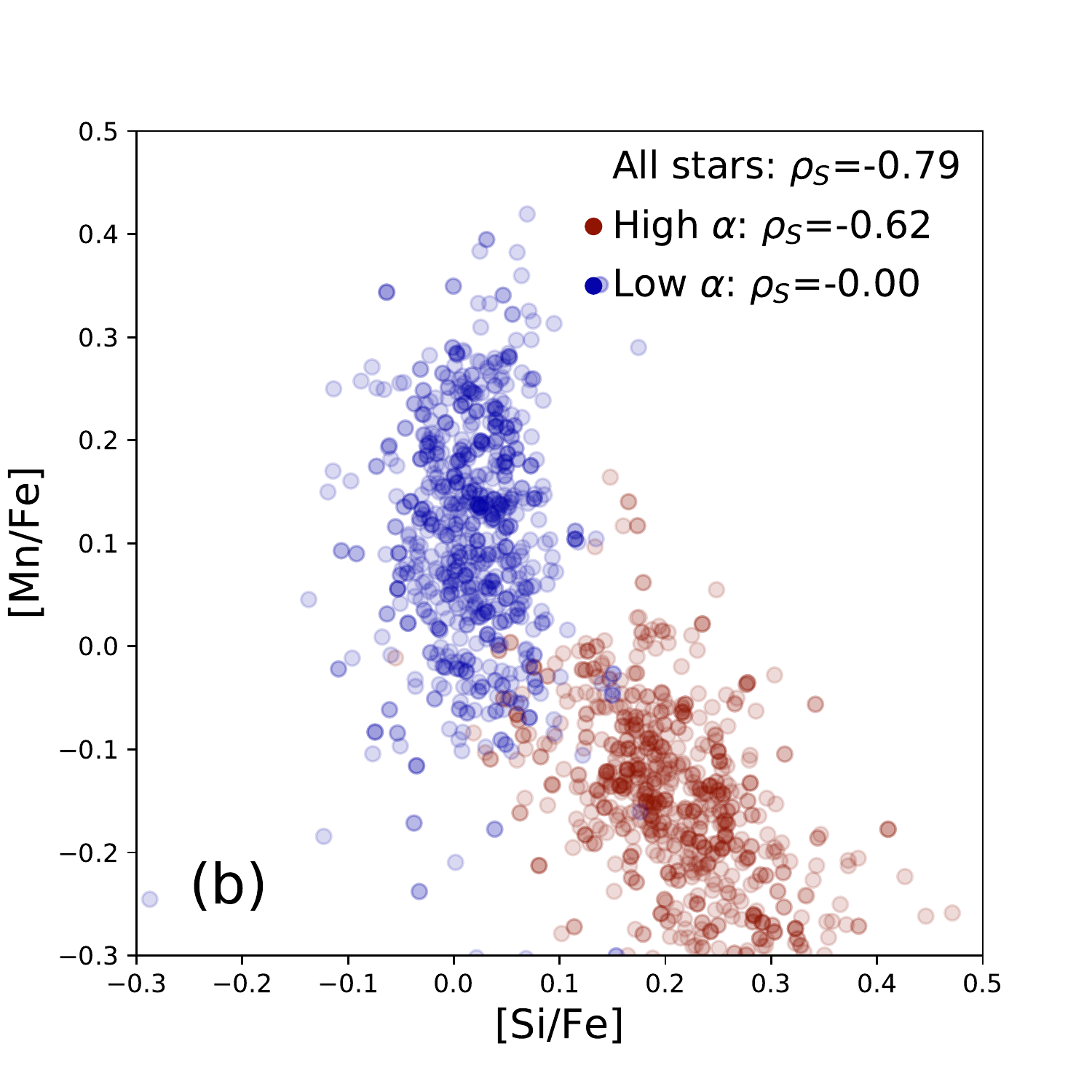} 
\caption{
Sample abundance-abundance correlations, highlighting (a) similarities and (b) differences between correlations in the high-$\alpha$ (red) and low-$\alpha$ (blue) abundance groups.  Values of $\rho_S$ approaching $+1$ and $-1$ indicate strong positive and negative linear correlation, respectively.
}
\label{fig:element_correlations_plots}
\end{center}
\end{figure*}

These correlations can also be calculated separately for high- and low-$\alpha$-enhancement stars (as in \S\ref{sec:sn_comparison}, ${\rm [\alpha/M] < +0.1}$ and ${\rm [\alpha/M] > +0.12}$), which is roughly equivalent to a metallicity separation at ${\rm [Fe/H] \sim -0.1}$.
Three main patterns emerge: 1) nearly identical $\rho_S$ and behavior in the [X$_1$/Fe]--[X$_2$/Fe] plane between the high- and low-$\alpha$ groups, 2) very similar $\rho_S$ values but offset trends in the [X$_1$/Fe]--[X$_2$/Fe] plane, and 3) very different $\rho_S$ values.  Exemplars of these latter two patterns are shown in Figure~\ref{fig:element_correlations_plots}, where stars have been restricted to those with $T_{\rm eff} \ge 3800$~K to reduce the systematic scatter described in \S\ref{sec:xfe_vs_feh}.

\vspace{-4pt}
\begin{enumerate} \itemsep -2pt
\item Frequently, pairs of elements with identical behavior in the [X$_1$/Fe]--[X$_2$/Fe] plane between the high- and low-$\alpha$ groups are those with high scatter.  

\item All pairs of $\alpha$-elements share behavior similar to [O/Fe]--[Si/Fe], shown in Figure~\ref{fig:element_correlations_plots}a.  The horizontal and vertical ``offsets'' here are expected from the bimodal distribution (by construction) of the $\alpha$-elements and their correlations; the very similar quantitative relationship between these elements is also expected due to their similar formation sites at all metallicities.

\item 
In contrast, Figure~\ref{fig:element_correlations_plots}b shows that the relationship between [Si/Fe] and [Mn/Fe] differs significantly between the high-$\alpha$ and low-$\alpha$ groups --- [Si/Fe] and [Mn/Fe] are not measurably correlated in the metal-rich, low-$\alpha$ population but are anti-correlated in the metal-poor population.  The lack of metal-rich correlation is driven by the flat trend in [Si/Fe] at the same supersolar metallicities where [Mn/Fe] continues to rise.  At lower metallicities, the anti-correlation reflects the fact that the CCSN Mn yields have a metallicity dependence while the Si ones do not.

\end{enumerate}

\begin{figure*}[!hptb]
\includegraphics[angle=0,trim=0in 0in 0in 0in, clip, width=0.95\textwidth]{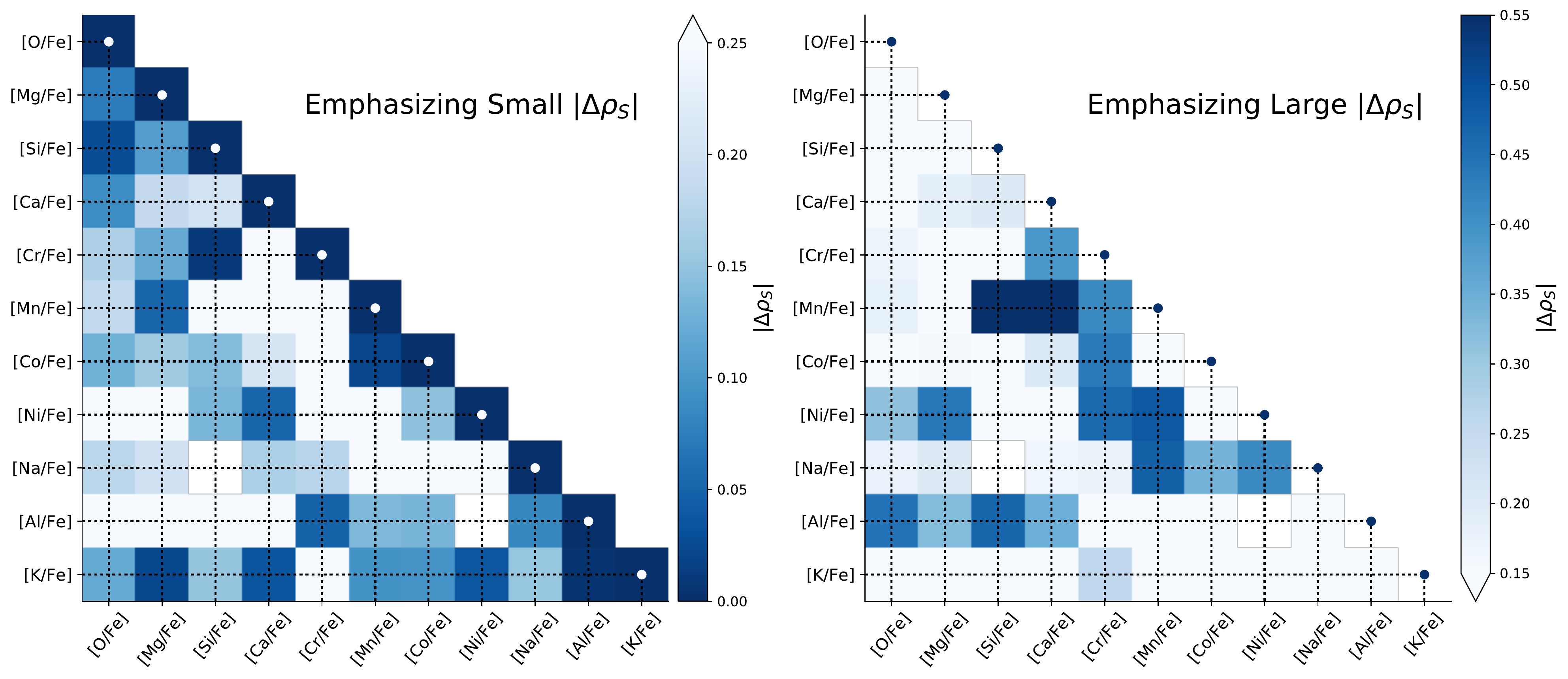}
\caption{
The absolute difference in correlation coefficient, $|\Delta\rho_S|$, in each pair of elements between the high- and low-$\alpha$ stars.  The colorbar in the left panel is scaled such that small differences are emphasized in dark blue; the colorbar in the right panel is scaled such that large differences are emphasized.
}
\label{fig:element_correlations_diff}
\end{figure*}

A graphical summary of the differences in all $\rho_S$(X$_1$,X$_2$) between the high- and low-$\alpha$ stars is shown in Figure~\ref{fig:element_correlations_diff}.  The values shown in both grids are identical, but the colors are scaled to emphasize small differences on the left and large differences on the right (as in Figure~\ref{fig:element_correlations}, pairs with $p$-values greater than 0.05 are blanked out). 
For example, in the right panel, [Mn/Fe] stands out as having different correlations in high- and low-$\alpha$ stars with elements whose correlation with [Fe/H] changes in different metallicity regimes (e.g., Ni) or whose trend is flat in the two bins considered here (e.g., Ca and Si).  In the left panel, [O/Fe], which is dominated by weakly metallicity-dependent CCSNe contributions, has a similar correlation at all $\alpha$ abundance with elements whose yields' metallicity dependence does not change significantly across the [Fe/H] range probed here.  The inclusion of Cr in this set, which has non-negligible production in SN~Ias, may indicate that SN~Ia [Cr/Fe] yields are metallicity-independent, as the CCSNe yields appear to be.

\section{Summary}
\label{sec:summary}

Using chemical abundances determined in $\sim$4000 APOGEE stars (SDSS DR14/DR15) towards the inner Milky Way, we have described the basic abundance patterns of 11 elements.  We found generally good agreement with patterns in the literature probing the same region of the Galaxy, though with some differences that may be attributable to ASPCAP issues or to differing sample sizes.  The position of the $\alpha$-element abundance knee or downturn, as measured in the [Mg/Fe]--[Fe/H] plane with two example metrics, was found to be nearly constant with $R_{\rm GC}$ and interpreted as evidence for a well-mixed ISM in the early MW and/or high levels of radial mixing post-star formation.  The narrow [Fe/H] span of the transition region from high- to low-$\alpha$ abundance (compared to farther out in the disk), and the lack of a bimodality in the [$\alpha$/Fe] distributions at subsolar metallicity, suggest that the abundances are dominated by a single chemical evolutionary sequence and do not reflect large amounts of mixing from regions with radically different enrichment histories.

The linear correlation between pairs of elements were found to vary in behavior when the sample is divided into high-$\alpha$-abundance and low-$\alpha$-abundance groups.  Some elemental pairs have very similar correlations in the high- and low-$\alpha$ groups (e.g., the $\alpha$-elements themselves), while other pairs differ between the groups (e.g., [Mg/Fe] vs [Ni/Fe]).  If interpreted as signatures of the varying impact of different nucleosynthetic pathways at different stages of the MW's evolution, or of the metallicity dependence of the nucleosynthetic yields themselves, these empirical observations provide important constraints on the chemical history of the inner Galaxy.  The wide range of metallicities probed, especially the poorly studied regime at ${\rm [Fe/H] > 0}$, and the seeming simplicity of the dominant enrichment history, render this part of the MW uniquely important for testing and refining our understanding of chemical evolutionary processes.

\begin{acknowledgments}
We thank the anonymous referee for very thoughtful and helpful comments that improved the clarity of this paper.
G.Z.\ acknowledges support from the Barry~M.~Lasker Data Science Research Fellowship, sponsored by the Space Telescope Science Institute in Baltimore, MD, USA. 
M.S.\ acknowledges the Programme National de Cosmologie et Galaxies (PNCG) of CNRS/INSU, France, for financial support.
T.C.B.\ acknowledges partial support from grant PHY 14-30152 (Physics Frontier Center/JINA-CEE), awarded by the U.S. National Science Foundation.
H.J.\ acknowledges support from the Crafoord Foundation, and Stiftelsen Olle Engkvist Byggm\"astare.

Funding for SDSS-III has been provided by the Alfred P. Sloan Foundation, the Participating Institutions, the National Science Foundation, and the U.S. Department of Energy Office of Science. The SDSS-III web site is \url{http://www.sdss3.org/}.

SDSS-III is managed by the Astrophysical Research Consortium for the Participating Institutions of the SDSS-III Collaboration including the University of Arizona, the Brazilian Participation Group, Brookhaven National Laboratory, Carnegie Mellon University, University of Florida, the French Participation Group, the German Participation Group, Harvard University, the Instituto de Astrof\'isica de Canarias, the Michigan State/Notre Dame/JINA Participation Group, Johns Hopkins University, Lawrence Berkeley National Laboratory, Max Planck Institute for Astrophysics, Max Planck Institute for Extraterrestrial Physics, New Mexico State University, New York University, The Ohio State University, Pennsylvania State University, University of Portsmouth, Princeton University, the Spanish Participation Group, University of Tokyo, University of Utah, Vanderbilt University, University of Virginia, University of Washington, and Yale University.

Funding for the Sloan Digital Sky Survey IV has been provided by the Alfred P. Sloan Foundation, the U.S. Department of Energy Office of Science, and the Participating Institutions. SDSS-IV acknowledges support and resources from the Center for High-Performance Computing at the University of Utah. The SDSS web site is \url{www.sdss.org}.

SDSS-IV is managed by the Astrophysical Research Consortium for the Participating Institutions of the SDSS Collaboration including the Brazilian Participation Group, the Carnegie Institution for Science, Carnegie Mellon University, the Chilean Participation Group, the French Participation Group, Harvard-Smithsonian Center for Astrophysics, Instituto de Astrof\'isica de Canarias, The Johns Hopkins University, Kavli Institute for the Physics and Mathematics of the Universe (IPMU) / University of Tokyo, Lawrence Berkeley National Laboratory, Leibniz Institut f\"ur Astrophysik Potsdam (AIP),  Max-Planck-Institut f\"ur Astronomie (MPIA Heidelberg), Max-Planck-Institut f\"ur Astrophysik (MPA Garching), Max-Planck-Institut f\"ur Extraterrestrische Physik (MPE), National Astronomical Observatories of China, New Mexico State University, New York University, University of Notre Dame, Observat\'ario Nacional / MCTI, The Ohio State University, Pennsylvania State University, Shanghai Astronomical Observatory, United Kingdom Participation Group, Universidad Nacional Aut\'onoma de M\'exico, University of Arizona, University of Colorado Boulder, University of Oxford, University of Portsmouth, University of Utah, University of Virginia, University of Washington, University of Wisconsin, Vanderbilt University, and Yale University.

We thank the E-Science and Supercomputing Group at Leibniz Institute for Astrophysics Potsdam (AIP) for their support with running the StarHorse code on AIP cluster resources.
\end{acknowledgments}

\bibliographystyle{yahapj}
\bibliography{ms}

\clearpage
\appendix
\section{Solar Radius Comparison Sample}
\label{sec:app_snsample}

\begin{figure*}
\begin{center}
\includegraphics[angle=0,trim=1in 0in 1in 0in, clip, width=0.85\textwidth]{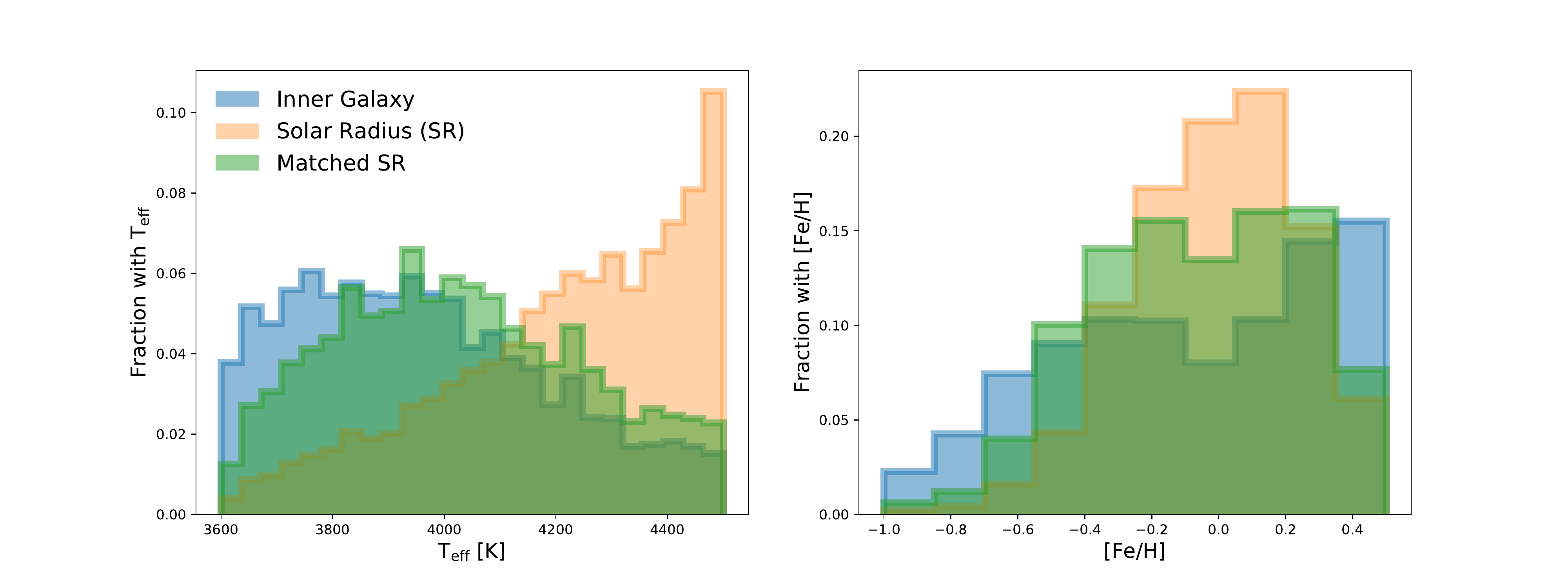}  
\caption{
Parameter distributions for the primary inner Galaxy sample (blue), the full set of SR stars meeting the bulge parameter and data quality limits (orange), and the subsample of SR stars (green) selected to have a  $T_{\rm eff}$--[Fe/H] joint distribution similar to the inner Galaxy sample.
}
\label{fig:matched_sn_params}
\end{center}
\end{figure*}

Figure~\ref{fig:matched_sn_params} shows the parameter distributions of the main bulge sample and the matched SR sample.  The full set of SR stars meeting the parameter and data quality limits described in \S\ref{sec:sel_criteria}, shown in orange, is (unsurprisingly) significantly more skewed towards warmer stars with a narrower range of metallicity than the bulge sample (blue).  The green distributions show the SR sample after downsampling in the joint $T_{\rm eff}$--[Fe/H] plane to match the bulge sample as closely as possible.  The metal-poor and coolest stars are the least represented in the SR sample, which is why the stars with $T_{\rm eff} < 3800$~K are not considered in the quantitative comparisons in \S\ref{sec:sn_comparison}.

The requirements for reliable stellar parameters and abundances (\S\ref{sec:sel_criteria}) help drive the average signal-to-noise ratios (SNR) of both the SR and inner Galaxy samples to higher than APOGEE's nominal goal of 100 per pixel.  The SR sample has higher SNR (median of 369) than the inner Galaxy sample (median of 148), but because both are comfortably above the threshold where noise has a significant impact, the typical abundance uncertainties are nearly identical between the samples (as seen, for example, in Figures~\ref{fig:abundances_grid} and \ref{fig:abundances_grid_sn}).

\section{``High-[O/Fe]'' Stars}
\label{sec:app_highO}

Table~\ref{tab:high_o_stars} contains the stars described in \S\ref{sec:alpha_elements} as having artificially enhanced $\alpha$ abundances, particularly [O/Fe] and [Ca/Fe].  We report two [$\alpha$/M] measurements for each star: one derived using the Kurucz grid of model atmospheres, which is the result reported as the global ALPHA\_M parameter in DR14/15 \citep{Holtzman_2018_dr13dr14apogee}, and one derived using the MARCS grid of model atmospheres.  As described in \S\ref{sec:alpha_elements}, we argue that the lower MARCS-based $\alpha$ measurements indicate the ``high-O'' abundances are due to poor fitting by the Kurucz atmosphere-based synthetic grid.

\begin{table*}[!hptb]
\begin{center}
\begin{tabular}{ccc}
 \hline
 \hline
\multirow{2}{*}{2MASS ID} & \multicolumn{2}{c}{[$\alpha$/M]} \\
 & Kurucz & MARCS \\
 \hline
2M17165161-2820586 & 0.19 & 0.12 \\
2M17165888-2647525 & 0.20 & 0.09 \\
2M17171732-2430268 & 0.20 & 0.13 \\
2M17175971-2515548 & 0.21 & 0.15 \\
2M17195372-2916228 & 0.20 & 0.14 \\
2M17345097-1940289 & 0.19 & 0.15 \\
2M17375797-2255290 & 0.20 & 0.12 \\
2M17425975-2727054 & 0.17 & 0.15 \\
2M17431651-2449057 & 0.17 & 0.15 \\
2M17463735-2707474 & 0.20 & 0.13 \\
2M17481951-2300243 & 0.20 & 0.05 \\
2M17500262-2247012 & 0.19 & 0.12 \\
2M17500582-2317042 & 0.19 & 0.11 \\
2M17503099-2252536 & 0.18 & 0.10 \\
2M17505103-2321525 & 0.21 & 0.14 \\
2M17540467-2138051 & 0.21 & 0.10 \\
2M17553603-2910288 & 0.18 & 0.12 \\
2M17570384-2057554 & 0.20 & 0.11 \\
2M18000976-2903162 & 0.20 & 0.14 \\
2M18011080-1808278 & 0.21 & 0.10 \\
2M18011227-1907015 & 0.15 & 0.10 \\
2M18022227-1712193 & 0.21 & 0.12 \\
2M18023639-2839312 & 0.21 & 0.17 \\
2M18032477-2156215 & 0.21 & 0.18 \\
2M18035010-1719038 & 0.21 & 0.10 \\
2M18043933-2502198 & 0.19 & 0.11 \\
2M18061670-1815549 & 0.19 & 0.11 \\
2M18090611-2436574 & 0.20 & 0.13 \\
2M18100202-0809009 & 0.20 & 0.11 \\
2M18113357-2706583 & 0.21 & 0.17 \\
2M18120561-2346546 & 0.19 & 0.13 \\
2M18120591-2749555 & 0.19 & 0.11 \\
2M18125005-2734185 & 0.20 & 0.13 \\
2M18264870-1517562 & 0.21 & 0.11 \\
2M18423748-3014180 & 0.19 & 0.11 \\
 \hline
\end{tabular}
\caption{
2MASS IDs and [$\alpha$/M] measurements for the stars described in \S\ref{sec:alpha_elements} as artificially ``high-O'' stars.
}
\label{tab:high_o_stars}
\end{center}
\end{table*}

\end{document}